\documentclass[a4paper,11pt]{article}
\usepackage{aaskaiid}

\usepackage{acronym}
\usepackage{wrapfig}

\acrodef{NS}[NS]{neutron star}
\acrodef{DM}[DM]{dark matter}
\acrodef{GR}[GR]{General Relativity}
\acrodef{EoS}[EoS]{equation of state}
\acrodef{QCD}[QCD]{quantum chromodynamics}
\acrodef{MoI}[MoI]{moment of inertia}
\acrodef{GW}[GW]{gravitational wave}
\acrodef{ToA}[ToA]{time of arrival}
\acrodef{PPM}[PPM]{Pulse Profile Modelling}
\acrodef{JBO}[JBO]{Jodrell Bank observatory}

\graphicspath{{./}{Figures/}}

\title{Probing Neutron Star Interiors and the Properties of Cold Ultra-dense Matter with the SKAO}
\ShortTitle{NS--EOS with the SKAO}

\author[1]{A. Basu}
\ShortName{A. Basu et al.}
\author[2]{V. Graber}
\author[3]{M. E. Lower}
\author[4]{M. Antonelli}
\author[1]{D. Antonopoulou}
\author[5,6]{M. Bagchi}
\author[7]{P. Char}
\author[8]{P. C. C. Freire}
\author[9,10,11]{B. Haskell}
\author[12,8]{H. Hu}
\author[13]{D. I. Jones}
\author[14]{B. Mukhopadhyay}
\author[15]{M. Oertel}
\author[16,17]{N. Rea}
\author[13]{V. Sagun}
\author[1]{B. Shaw}
\author[18]{J. Singha}
\author[1]{B. W. Stappers}
\author[19,1]{T. Thongmeearkom}
\author[20]{A. L. Watts}
\author[1]{P. Weltevrede}
\author[]{The SKA Pulsar Science Working Group}

\affiliation[1]{Jodrell Bank Centre for Astrophysics, Department of Physics and Astronomy, The University of Manchester, Manchester M13 9PL, UK}
\affiliation[2]{Department of Physics, Royal Holloway, University of London, Egham, TW20 0EX, UK}
\affiliation[3]{Centre for Astrophysics and Supercomputing, Swinburne University of Technology, PO Box 218, Hawthorn VIC 3122, Australia}
\affiliation[4]{CNRS/in2p3, Laboratoire de Physique Corpusculaire (LPC Caen), 14050 Caen, France}
\affiliation[5]{The Institute of Mathematical Sciences, Taramani, Chennai 600113, India}
\affiliation[6]{Homi Bhabha National Institute, Training School Complex, Anushakti Nagar, Mumbai 400094, India}
\affiliation[7]{Departamento de F\'isica Fundamental and IUFFyM, Universidad de Salamanca, Plaza de la Merced S/N, E-37008 Salamanca, Spain}
\affiliation[8]{Max-Planck-Institut f\"ur Radioastronomie, Auf dem H\"ugel 69, D-53121 Bonn, Germany}
\affiliation[9]{Dipartimento di Fisica, Universit\`{a} degli Studi di Milano, Via Celoria 16, 20133, Milano, Italy}
\affiliation[10]{INFN, Sezione di Milano, Via Celoria 16, 20133, Milano, Italy}
\affiliation[11]{Nicolaus Copernicus Astronomical Center of the Polish Academy of Sciences, Bartycka 18, 00-716, Warsaw, Poland}
\affiliation[12]{Lohrmann Observatory, Technische Universit\"at Dresden, Mommsenstraße 13, 01062 Dresden, Germany}
\affiliation[13]{Mathematical Sciences and STAG Research Centre, University of Southampton, Southampton SO17 1BJ, United Kingdom}
\affiliation[14]{Department of Physics, Indian Institute of Science, C. V. Raman Road, Bangalore 560012, India}
\affiliation[15]{Observatoire astronomique de Strasbourg, CNRS, Universit\'e de Strasbourg, 11 rue de l'Universit\'e, 67000 Strasbourg, France}
\affiliation[16]{Institute of Space Sciences (ICE-CSIC), Campus UAB, C/ de Can Magrans s/n, Cerdanyola del Vallès (Barcelona) 08193, Spain}
\affiliation[17]{Institut d'Estudis Espacials de Catalunya (IEEC), Castelldefels, Spain}
\affiliation[18]{High Energy Physics, Cosmology \& Astrophysics Theory (HEPCAT) Group, Department of Mathematics and Applied Mathematics, University of Cape Town, Cape Town 7700, South Africa}
\affiliation[19]{National Astronomical Research Institute of Thailand, Don Kaeo, Mae Rim, Chiang Mai 50180, Thailand}
\affiliation[20]{Anton Pannekoek Institute for Astronomy, University of Amsterdam, Science Park 904, 1098XH Amsterdam, the Netherlands}

\emailAdd{avishek.basu@manchester.ac.uk}
\emailAdd{Vanessa.Graber@rhul.ac.uk}
\emailAdd{mlower@swin.edu.au}

\abstract{
Matter inside neutron stars is compressed to densities several times greater than nuclear saturation density, while maintaining low temperatures and large asymmetries between neutrons and protons. Neutron stars, therefore, provide a unique laboratory for testing physics in environments that cannot be recreated on Earth. To uncover the highly uncertain nature of cold, ultra-dense matter, discovering and monitoring pulsars is essential, and SKAO will play a crucial role in this endeavour. In this chapter, we will present the current state-of-the-art in dense matter physics and dense matter superfluidity, and discuss recent advances in measuring global neutron star properties (masses, moments of inertia, and maximum rotation frequencies) as well as non-global observables (pulsar glitches and free precession). We will specifically highlight how radio observations of isolated neutron stars and those in binaries---such as those performed with SKAO in the near future---inform our understanding of ultra-dense physics and address in detail how SKAO's telescopes unprecedented sensitivity, large-scale survey and sub-arraying capabilities will enable novel dense matter constraints. We will also address the potential impact of dark matter and modified gravity models on these constraints and emphasise the role of synergies between SKAO and other facilities, specifically X-ray telescopes and next-generation gravitational wave observatories.}

\begin{document}
\maketitle

\section{Introduction}

\begin{figure*}
\centering
\includegraphics[width=0.9\linewidth]{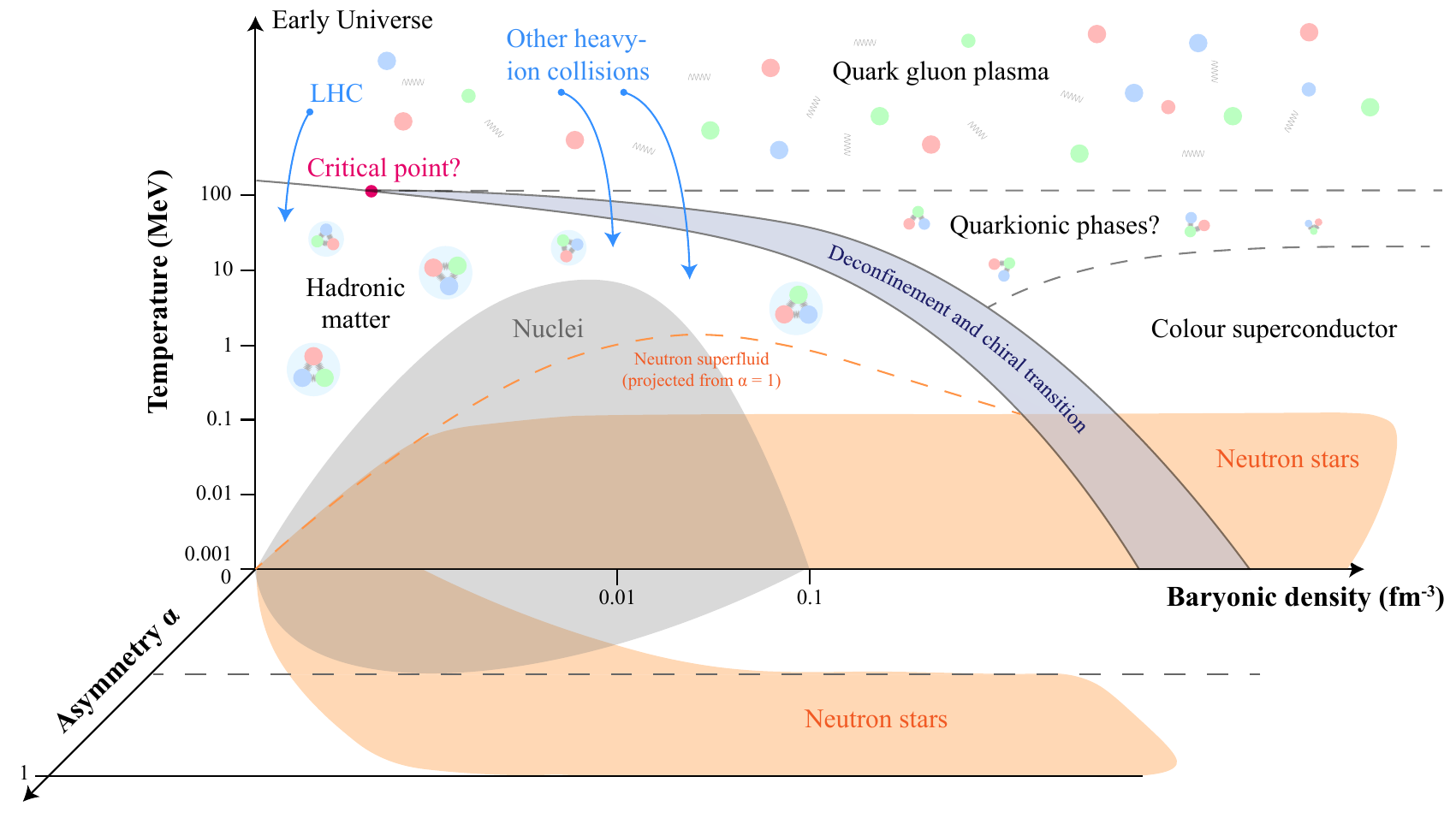}
\caption{The parameter space and states of matter present in \acp{NS}, as compared to terrestrial experiments. The figure shows temperature against baryon number density against asymmetry, $\alpha = 1 - 2 Y_q$, where $Y_q$ is the hadronic charge fraction (generally equal to the ratio of the proton number to the total number of baryons). $\alpha = 0$ for matter with equal numbers of neutrons and protons, and $\alpha = 1$ for pure neutron matter. The orange regions show the parameter space occupied by \acp{NS} projected onto the temperature-baryon density and asymmetry-baryon density planes, respectively. The grey regions show projections onto the same planes for isolated nuclei, which exist up to $\alpha \approx 0.3$. Above this value, one would find a mix of nuclei and light particles.}
\label{fig:trho}
\end{figure*}

Formed in the core-collapse supernovae of massive stars, \acf{NS} interiors are governed by ultra-high densities, low temperatures and large asymmetries in proton/neutron number. Understanding the nature and properties of matter under such extreme conditions is one of the key unsolved challenges in modern science. Because these environments cannot be probed in terrestrial experiments, \acp{NS} are the only laboratories to study matter under high-density, low-temperature and large neutron/proton number asymmetry conditions (see Figure~\ref{fig:trho}) and, thus, advance our general understanding of nuclear physics and \acf{QCD}~\citep{Chatziioannou:2024tjq}. This chapter will outline the fundamental role that SKAO will play in this endeavour.

\begin{figure*}
\centering
\includegraphics[width=0.9\linewidth]{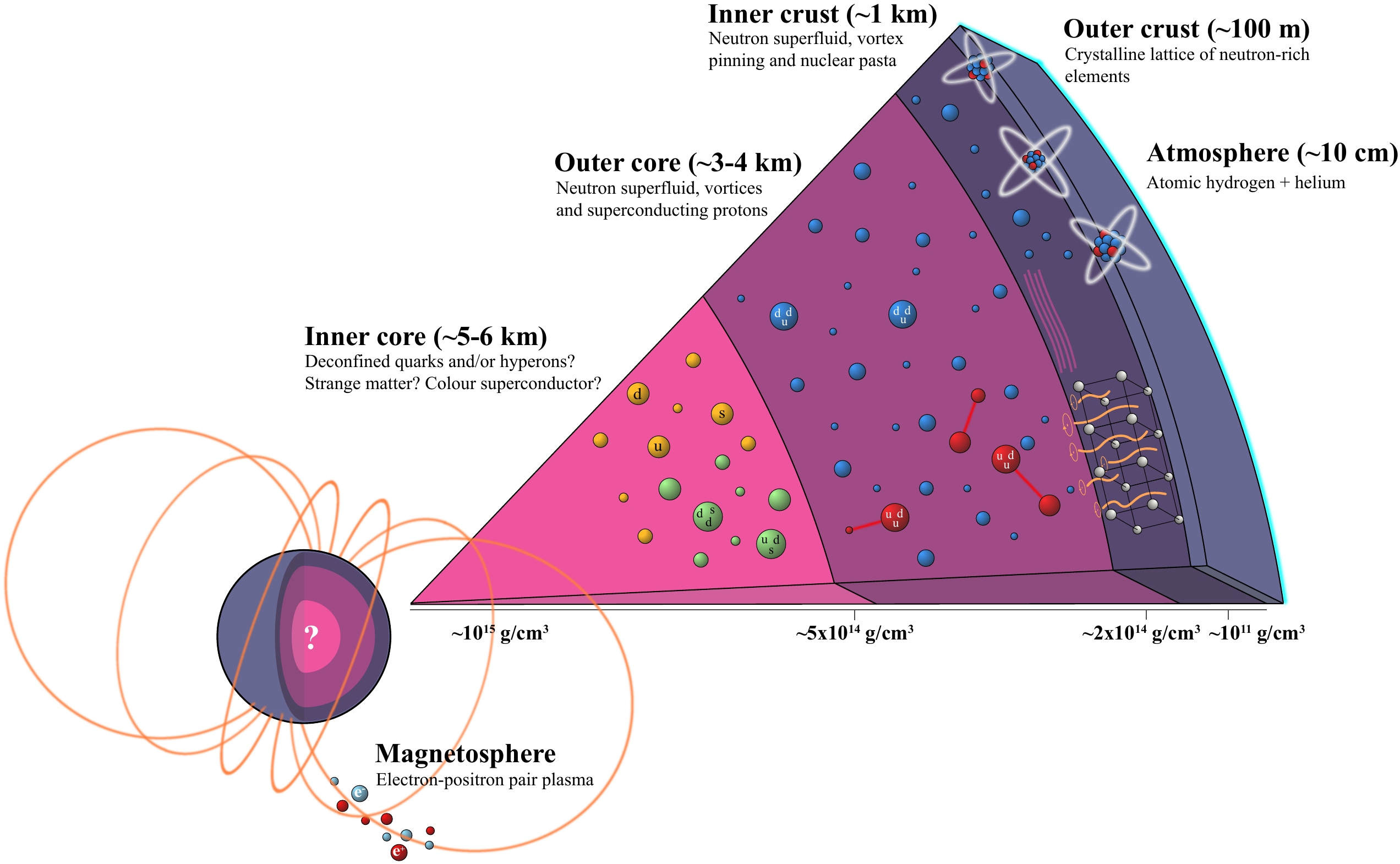}
\caption{Schematic structure of a \ac{NS}: The outer layer---a solid crust of fully ionised nuclei---is supported mainly by electron degeneracy pressure. The inner crust starts around the neutron drip density, $4 \times 10^{11} \, \text{g/cm}^3$, where neutrons begin to leak out of the nuclei. From this point on, neutron degeneracy pressure starts to contribute. At densities of approximately $2 \times 10^{14} \, \text{g/cm}^3$, at the crust-core boundary, nuclei dissolve entirely. In the core, densities can reach up to ten times the nuclear saturation density---the typical density of atomic nuclei---and the pressure due to the repulsive channels of the nuclear interaction is essential to counterbalance gravity.}
\vspace{-10pt}
\label{fig:schema}
\end{figure*}

A plethora of physical parameters and processes affect the properties of \ac{NS} interiors. Due to their enormous gravitational forces, internal \ac{NS} densities vary by many orders of magnitude resulting in dramatic composition changes across the star (see Figure~\ref{fig:schema}), even though the exact locations of these transitions remain unknown. The primary macroscopic diagnostic to address open questions about the \ac{NS} structure and dense matter interactions is the pressure-density-temperature relation of bulk matter, the \acf{EoS}. Constraining the \ac{EoS} is, thus, essential to inferring key aspects of dense matter microphysics. In addition to uncertainties in composition, we also do not know how ultra-dense \ac{NS} matter behaves dynamically; a question which cannot be answered through \ac{EoS} constraints alone. The key aspect here is that, although \acp{NS} are born hot, they cool down to temperatures well below nuclear energy scales---around $10^9\, {\rm K}$---within months to years~\citep{Page_etal2004}. As a result, \acp{NS} also exhibit the rich phenomenology of low-temperature systems, with at least three distinct macroscopic quantum phases occupying the \ac{NS} interior~\citep{Chamel2017}.

Uncertain composition and dynamical properties are, however, not only of interest from a dense matter perspective. Both play a critical role in astrophysics, with the \ac{EoS} influencing, for example, the dynamics of binary \ac{NS} mergers and the corresponding \acf{GW} signal and nucleosynthesis~\citep{De_etal2018, LIGOScientific:2018hze} as well as core collapse supernovae and the associated neutrino signal~\citep{Janka_etal2007}. Amongst other things, superfluidity (together with the presence of strong magnetic fields) affects \ac{NS} cooling~\citep{Page_etal2004}, the star's rotational evolution in the form of pulsar glitches~\citep{Haskell15,Antonopoulou:2022rpq,
Zhou:2022,Antonelli:2023vpd}, and internal oscillations, particularly relevant for next-generation \ac{GW} facilities~\citep{Andersson:2001kx}. Understanding dense matter is, thus, crucial to understanding numerous high-energy astrophysical phenomena, many of which are multi-messenger events, and SKAO's radio observations will make essential contributions to this.

The paper is organised as follows. Sections~\ref{sec:unknowns_dmphysics} and~\ref{sec:unknowns_SFphysics} summarise the current state-of-the-art and existing challenges in dense matter physics and dense matter superfluidity, respectively. Section~\ref{sec:radioPSR_NP_connection} gives an outline of how observations of radio pulsars constrain nuclear physics properties, highlighting five distinct classes of \ac{NS} observables and existing nuclear physics constraints. The potential impact of dark matter and modified theories of gravity on these insights is discussed in Section~\ref{sec:DM_modgrav}. We then outline the crucial role that SKAO will play in constraining dense matter uncertainties in the coming decades in Section~\ref{sec:SKA_expect}. We will specifically discuss the observing modes that are required to unlock SKAO's full potential to uncover unknown nuclear physics, while also addressing synergies with other astronomical facilities. Conclusions are presented in Section~\ref{sec:conclusions}.

\subsection{Unknowns in dense matter physics}
\label{sec:unknowns_dmphysics}

The dense matter composing \acp{NS} is governed by the strong interaction in a regime of densities, temperatures, and asymmetries (neutron to proton ratio) neither accessible by terrestrial experiments nor by ab-initio theoretical calculations. Our current understanding of \ac{NS} interiors summarised in Figure~\ref{fig:schema} invokes an outer solid crust with a Coulomb lattice composed of nuclear ions and an electron gas as the outermost region of the star. Moreover, the core is composed of homogeneous strongly interacting matter permeated by an ideal gas of electrons and muons required by electrical charge neutrality.

Only the outer crustal layers contain nuclei for which masses can be measured~\citep{Baym:1971pw,Huang:2021nwk}. Deeper in the crust, descriptions of the properties of neutron-rich nuclei and the neutron fluid close to the crust-core boundary rely on theoretical calculations~\citep{Baym:1971ax,Grill:2014aea,Pearson:2018tkr}. Although model-dependent to some extent, measured properties of finite nuclei and ab-initio approaches to low density neutron matter allow us to sufficiently infer the composition of the inner crust~\citep{Douchin:2001sv,Potekhin:2013qqa}. However, the superfluid and transport properties are more difficult to constrain (see below). For a detailed discussion of the physics of the \ac{NS} crust see, e.g.,~\cite{ChamelHaensel2008} and references therein. When modelling the \ac{NS} core, ab-initio approaches attempt to solve the nuclear many-body problem using few-body interactions as a starting point. While nucleonic two-body interactions at low energies are well constrained by experiments, three- and more-body forces are still a frontier in nuclear physics with some progress at low energies in the last decades due to the development of interactions based on effective field theories~\citep{Tews:2012fj,Hebeler:2013nza,Lynn:2015jua,Drischler:2017wtt,Huth:2021bsp}. Additional uncertainties arise from solving the many-body problem based on different methods, which all include approximations. More phenomenological approaches are based on energy density functionals with parameters determined by nuclear data, benchmark ab-initio calculations, or astrophysical data.

At the high densities reached in \ac{NS} cores, there might be transitions to non-nucleonic states of matter~\citep{Glendenning:1998ag,HeiselbergHjorth-Jensen2000}. Possibilities include the formation of hyperons (strange baryons), $\Delta$-baryons, pion or kaon condensates~\citep{Tolos:2020aln}, or quark matter (leading to the formation of so-called \textit{hybrid} stars), possibly in colour-superconducting states~\citep{Alford:2007xm}. Under some conditions, such hybrid stars exhibit a twin phenomenon: two stars that have approximately the same mass but significantly different radii~\citep{Glendenning:1998ag, Blaschke:2019tbh}. It is even possible that the entire star converts into a lower energy self-bound state consisting of up, down and strange quarks, known as a strange quark star~\citep{Witten:1984rs,Weber:2004kj}. The densities at which such phases would appear, and the degree to which they might co-exist with other phases, are highly uncertain~\citep{Buballa:2014jta,Annala:2019puf, Constantinou:2023ged,Essick:2023fso}. For the non-nucleonic degrees of freedom in the \ac{NS} core, experimental information is scarce or non-existent and uncertainties are much larger than for purely nuclear systems. For reviews on different approaches to describe dense nuclear and \ac{NS} matter see, e.g.,~\citet{Oertel:2016bki,Burgio:2021vgk}. 

Existing insights into the \ac{NS} \ac{EoS} comprise theoretical benchmark calculations and, in particular, chiral effective field theory calculations for low-density neutron matter up to around $1.5$ the nuclear saturation density, $\rho_0$~\citep{Keller:2022crb}, and perturbative QCD calculations at very high densities $(\gtrsim 40 \rho_0)$~\citep{Gorda:2022jvk}. Experimentally, \ac{EoS} constraints arise from data on nuclear masses, nuclear resonances, polarisabilities, heavy-ion collisions, and neutron skin thickness measurements with a recent update by PREX-II and CREX~\citep{PREX:2021umo,CREX:2022kgg}.  \cite{MUSES:2023hyz} present a recent summary of these constraints, but see also \cite{Lattimer:2012xj,Oertel:2016bki} and references therein. Moreover, nuclear physics theory and experiments mostly concern low densities either close to symmetric or pure neutron matter, while \ac{NS} observations probe the neutron-rich (hence highly asymmetric) and high-density matter in \ac{NS} cores. 

From a modelling perspective, Bayesian analysis techniques have gained popularity as a systematic approach to infer the \ac{EoS} from available data. These are either based on a completely uninformed parametric~\citep{Read:2008iy,Lindblom:2012zi} or non-parametric~\citep{Landry:2018prl} representation of the \ac{EoS}, or on nuclear metamodelling techniques~\citep{Margueron:2017eqc,Char:2023fue,Scurto:2024ekq}. Although the crust does not have a significant impact on global \ac{NS} properties like mass $M$, radius $R$ or \ac{MoI} $I$ except for very low-mass \acp{NS}, it is important to note that only the construction of a so-called `unified' \ac{EoS}\footnote{A unified \ac{NS} \ac{EoS} requires a consistent determination of the core and crust \ac{EoS} and a thermodynamically consistent crust-core transition from the same underlying nuclear model.} ensures quantitatively reliable predictions for these global \ac{NS} properties~\citep{Fortin:2016hny,Suleiman:2021hre} and, therefore, a correct inference of \ac{EoS} properties from data. Moreover, the crust properties are important for interpreting other observables such as pulsar glitches (see below) and magnetar quasi-periodic oscillations~\citep[e.g.,][]{Gabler2018}. Publicly available tools such as CUTER~\citep{Davis:2024nda} allow us to complement any available core \ac{EoS} with a consistent crust.

Future astrophysical data will precisely determine the \ac{NS} \ac{EoS}. However, this may not be sufficient to unambiguously ascertain the composition of matter at the centre of the most massive \acp{NS} in the absence of a phase transition~\citep{Mondal:2021vzt,Xie:2020tdo,Iacovelli:2023nbv,Imam:2023ngm,Char:2025zdy}. This is due to a degeneracy, which arises from different compositions or nuclear interactions leading to the same $\beta$-equilibrated \ac{EoS}. Additional information, e.g., from \ac{NS} phenomena sensitive to transport properties such as cooling rates, pulsar glitches, or oscillation modes, will probably be necessary to fully explore the \ac{NS} interior structure.

We finally note that any inference of \ac{NS} properties from observational data is based on an underlying theory of gravity. While \ac{GR} is typically assumed to hold, modified theories of gravity affect the \ac{NS} structure and induce degeneracies when inferring dense matter properties~\citep{Danchev:2020zwn}. The same holds for the presence of particles beyond the standard model, such as dark matter candidates within \acp{NS}~\citep{Sandin:2008db, Dengler:2021qcq, Sagun:2022ezx}. To fully benefit from upcoming astrophysical data in informing dense subatomic physics, further research is required to reliably disentangle the role of such effects. We briefly return to these issues in Section~\ref{sec:DM_modgrav}.


\subsection{Unknowns in dense matter superfluidity}
\label{sec:unknowns_SFphysics}

Besides their extreme densities, the gravitational confinement renders \acp{NS} sufficiently long-lived for a weak $\beta$-equilibrium to be achieved. As a result, these compact objects can be considered cold, occupying the unique parameter space shown in Figure~\ref{fig:trho}. In fact, \acp{NS} older than a few hundred years are sufficiently cold for the nucleons to form Cooper pairs (analogous to terrestrial electronic superconductors), resulting in the appearance of distinct macroscopic quantum states~\citep{Baym1969, Sauls1989}. In particular, the free neutrons surrounding the lattice nuclei in the inner crust are superfluid, as is the neutron component in the outer core. The latter coexists with a condensate of superconducting protons. Moreover, additional superfluid phases, such as colour superconducting quark phases, might exist in the inner core~\citep{Alford:2007xm, Sedrakian:2018ydt}. 

Superfluid components alter the long-term evolution and dynamics of \acp{NS}. For example, superfluidity affects the star's thermal properties by suppressing nuclear reactions that cool it. Superfluidity also reduces the heat capacity but at the same time opens up new channels for neutrino emission, which can lead to faster cooling~\citep{Page_etal2004}. The corresponding net enhancement of cooling has been invoked to explain the behaviour of the young \ac{NS} in the Cassiopeia A supernova remnant, whose surface temperature may be decreasing faster than expected in standard cooling models due to the recent onset of neutron superfluidity~\citep{Shternin2011, Yakovlev2011}. Even though underlying temperature measurements are challenging~\citep{Posselt2022}, a confirmation of this anomaly would constrain the critical temperature for core superfluidity. Such observations would also place qualitative constraints on the superconducting transition temperature, as enhanced cooling requires this quantum phase to form well before the core superfluid~\citep{Shternin2021}. 

Such constraints are very valuable as transition temperatures directly relate to the highly uncertain superfluid pairing gaps (half the energy required to separate a Cooper pair)~\citep{Baldo1992}. The gaps are calculated assuming that the crustal neutrons and core protons pair in a state with zero spin and angular momentum (isotropic spin-singlet or $s$-wave pairing), while the core neutrons form Cooper pairs of non-zero spin and angular momentum (anisotropic spin-triplet or $p$-wave pairing)~\citep{Gezerlis2014}. However, calculations are complicated by in-medium effects, unknown many-body interactions above saturation, and anisotropy for the core neutrons. Moreover, superfluid energy gaps are typically calculated based on microscopic forces that differ from those considered for the composition and, subsequently, the \ac{EoS}, leading to inconsistencies in modelling superfluid properties. However, chiral effective field theory has recently enabled more consistent approaches for the singlet paired condensates~\citep{Lim2021}.


Superfluidity also affects stellar dynamics because superfluid components can flow relative to the normal fluid, increasing the system’s degrees of freedom~\citep{Glampedakis2011}. These microscopic properties manifest as observable phenomena, including sudden spin-up glitches in young \acp{NS}~\citep{espinoza2011, Basu:2021pyd} and timing noise across the pulsar population~\citep{Hobbs2010}. Neutron superfluids rotate by forming quantised vortices, whose interactions with their surroundings govern stellar rotation~\citep{Pines1980, Andersson2006}. In the canonical glitch model, vortices pin to the crustal lattice, decoupling part of the interior and storing angular momentum~\citep{Anderson1975, Pizzochero2011}. An avalanche of unpinning then rapidly recouples the components, producing the glitch. The shape of the observed signature, typically described in a body-averaged multi-component framework~\citep{haskell2015IJMPD}, reflects the frictional forces, unpinning and repinning processes, and the underlying nuclear physics.

In the inner crust, friction arises primarily from Kelvin wave excitation along vortex lines~\citep{Epstein1992, Jones1992, Graber2018ApJ}, while in the core it is dominated by electron scattering off vortex magnetic fields~\citep{Alpar1984b, Andersson2006}. Vortex magnetisation depends on entrainment—the non-dissipative coupling between nucleons—which also occurs in the crust via Bragg scattering~\citep{Chamel2012}. Although its strength is uncertain~\citep{Martin2016, Sauls2020}, entrainment critically determines the angular momentum available for glitches~\citep{andersson2012PhRvL, chamel2013PhRvL}. Magnetised vortices may also interact with proton fluxtubes, linking rotational and magnetic evolution~\citep{Ruderman1998, Sidery2009}; strong coupling could lead to vortex–fluxtube pinning and suppress slow precession~\citep{link_06}. The nature of proton superconductivity adds further uncertainty: while flux is often assumed to reside in a regular fluxtube array, it may be inhomogeneously distributed due to strong proton–neutron coupling~\citep{Wood2022}. Additional complexity may arise from colour-superconducting phases~\citep{Alford2010, Haber2018} and vortex or fluxtube tangling~\citep{Andersson2007}, whose dynamical impact remains poorly understood~\citep{haskell20}.

While theoretical progress in understanding the complicated processes outlined above remains essential, designated observing campaigns with the SKAO will provide an unparalleled view into dense matter properties beyond standard \ac{EoS} constraints~\citep{HaskellSedrakian}. As discussed below, high-precision timing with the SKAO's telescopes offers a unique way to probe related physics, surpassing existing constraints on the superfluid energy gaps from cooling observations and rough estimates of the superfluid \ac{MoI} from glitch observations.

\section{Connecting pulsar observations to dense matter physics: state-of-the-art}
\label{sec:radioPSR_NP_connection}

\subsection{General concepts}
\label{sec:general_concepts}
The TOV equations~\citep{Tolman1939, Oppenheimer1939} describe the equilibrium structure of \acp{NS}, linking macroscopic observables such as $M$, $R$, and $I$ to the dense matter \ac{EoS}. In the zero-temperature limit, the \ac{EoS} connects pressure and density based on nuclear microphysics, with corresponding uncertainties (see Section~\ref{sec:unknowns_dmphysics}) producing a range of models that map onto the $M$--$R$ or $M$--$I$ planes~\citep{Lindblom1992, Lattimer2005}. Measurements of \ac{NS} observables can therefore test different \ac{EoS} models. While direct radius measurements are infeasible in the radio, masses and rotation frequencies are readily accessible.
The primary technique enabling these measurements is high-precision pulsar timing~\citep{Hobbs2006}, which tracks every stellar rotation over long timescales. Comparing pulse times of arrival (\acsp{ToA}) with timing models yields precise rotational, astrometric, and—if applicable—orbital parameters. For recycled pulsars in binaries, five Keplerian parameters are routinely determined~\citep{Lorimer2008}, and for some systems, post-Keplerian parameters provide mass and potentially future $I$ constraints~\citep{Kramer:2021jcw}.

For isolated, slowly rotating pulsars, timing instead probes deviations from a smooth spin-down model. High-cadence monitoring reveals sudden spin-up glitches~\citep{espinoza2011} and correlated changes in emission, pulse profiles, and spin-down rates~\citep[][see also~\citealt{Oswald01.2026.SKA}]{Lyne2010, Brook+2016pks, Shaw:2022mxv, Basu:2024qwt, lower_etal_25}, interpreted as signatures of free precession~\citep{jones_andersson_02}. Such phenomena offer complementary insights into both global and internal \ac{NS} physics.

In the following, we summarise recent progress in measuring \ac{NS} masses, moments of inertia, and spin limits, along with studies of glitches and precession. We also outline how the SKAO with different Array Assembly (AA) configurations  will advance dense matter constraints and improve our understanding of \ac{NS} interiors (see Section~\ref{sec:SKA_expect}).


\subsection{Pulsar mass measurements}
\label{sec:mass}

\begin{figure*}[t!]
\centering
\includegraphics[width=0.95\linewidth]{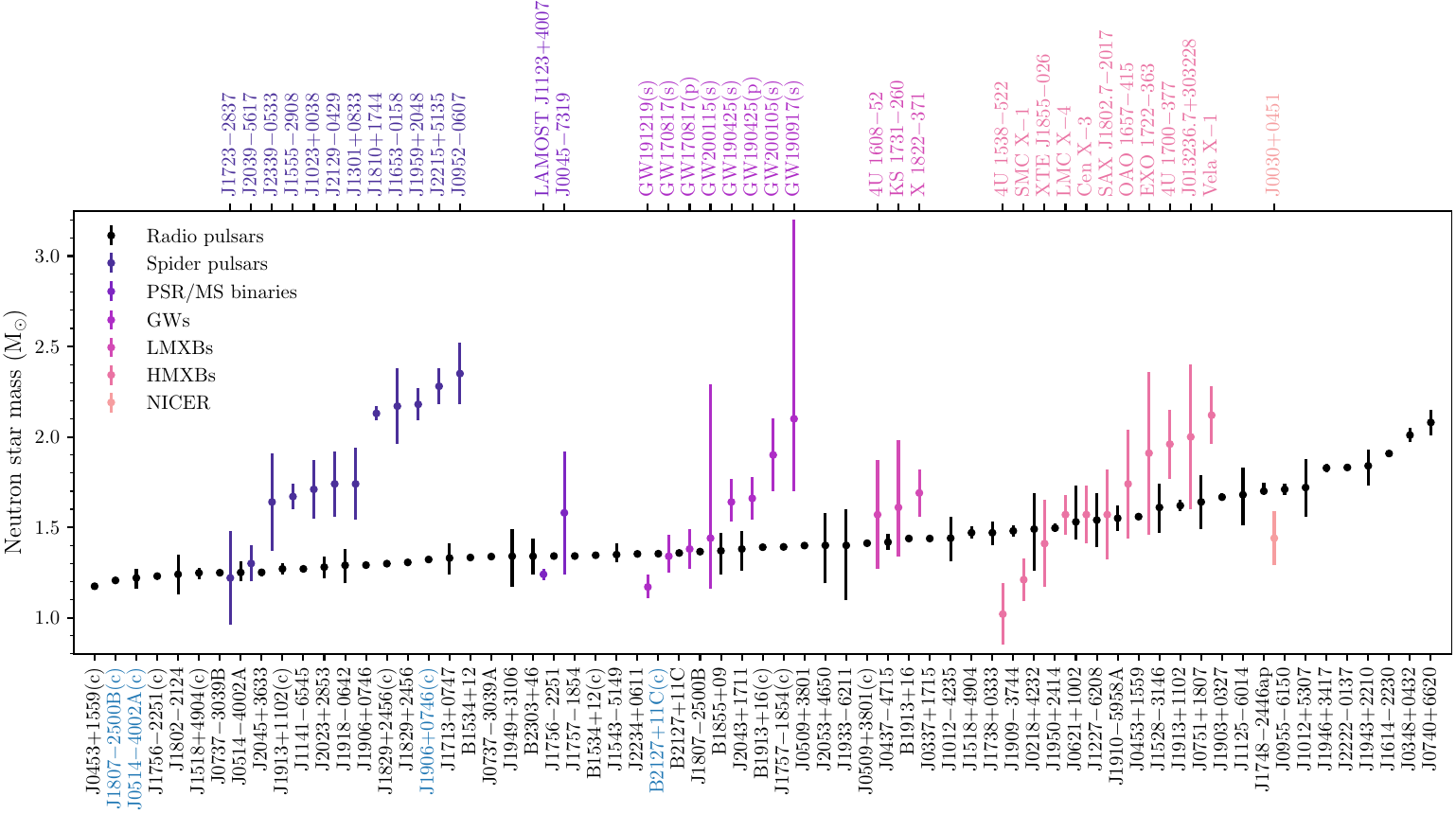}
\caption{The observed \ac{NS} mass spectrum with 68\% confidence intervals sourced from Extended Data Tables 1 and 2 of \cite{You2025} plus updated radio timing measurements (see \url{https://www3.mpifr-bonn.mpg.de/staff/pfreire/NS_masses.html}). The latter are shown in black. Other data points are redback and black widow systems (spider pulsars; dark purple), observations of pulsars with main-sequence companions (PSR/MS binaries; purple), gravitational wave events (GWs; light purple), low-mass X-ray binaries (LMXBs; magenta), high-mass X-ray binaries (HMXBs; pink) and NICER X-ray pulse profile modelling results (light pink). \acp{NS} indexed with a `(c)' indicate the companion to an observed pulsar. Note that those with names in blue could be either a \ac{NS} or a white dwarf based on current constraints. For \acp{NS} detected via \ac{GW} merger events, `(p)' indicates the primary or heavier object and `(s)' the secondary or less massive object.}
\label{fig:NS_masses}
\end{figure*}


To first post-Newtonian order, Einstein's equations governing relativistic orbital motion and radio-wave propagation in pulsar timing depend only on the component masses and observable Keplerian parameters~\citep{Damour1992}. The rotational stability of millisecond pulsars enables precise measurement of post-Keplerian parameters, like, the periastron advance $\dot{\omega}$, orbital decay from \ac{GW} emission $\dot{P}_{\rm b}$, time dilation and gravitational redshift $\gamma$, and the Shapiro delay parameters $r$ and $s$. Consequently, radio timing of binary pulsars (and one known stellar triple; \citealt{Ransom2014}), as planned with the SKAO's telescopes, provides our main means of determining spin frequencies~\citep[e.g.,][]{Basu:2021pyd} and accurate masses of \acp{NS} and white dwarfs~\citep[e.g.,][]{Guo2021, ColomiBernadich:2024A&A}, while also enabling stringent gravity tests~\citep[see][and
~\citealt{VenkatramanKrishnan01.2026.SKA}]{FreireWex2024}.
Future SKAO VLBI observations will further refine parallaxes, improving distance-dependent corrections to timing parameters and thereby enhancing mass and gravity constraints~\citep{Kramer:2021jcw}.

Over the last 15 years, radio pulsar timing has dramatically broadened the distribution of \ac{NS} masses. 61 corresponding mass measurements with a relative 1$\sigma$ uncertainty of less than 15\% are shown in black in Figure~\ref{fig:NS_masses}. We note that some of these pulsar-timing based measurements have been further enhanced through observations of diffractive scintillation induced by the scattering of radio waves in the interstellar medium (ISM; \citealt{Rickett1990}). The orbital motion of a pulsar can be imprinted as variations in the measured scintillation timescale, which allows us to uniquely identify the sense of the orbital inclination and further refine the component masses (see, \citealt{Reardon:2019MNRAS, Reardon:2020ApJ}).

\begin{figure*}[t!]
\centering
\includegraphics[width=0.8\linewidth]{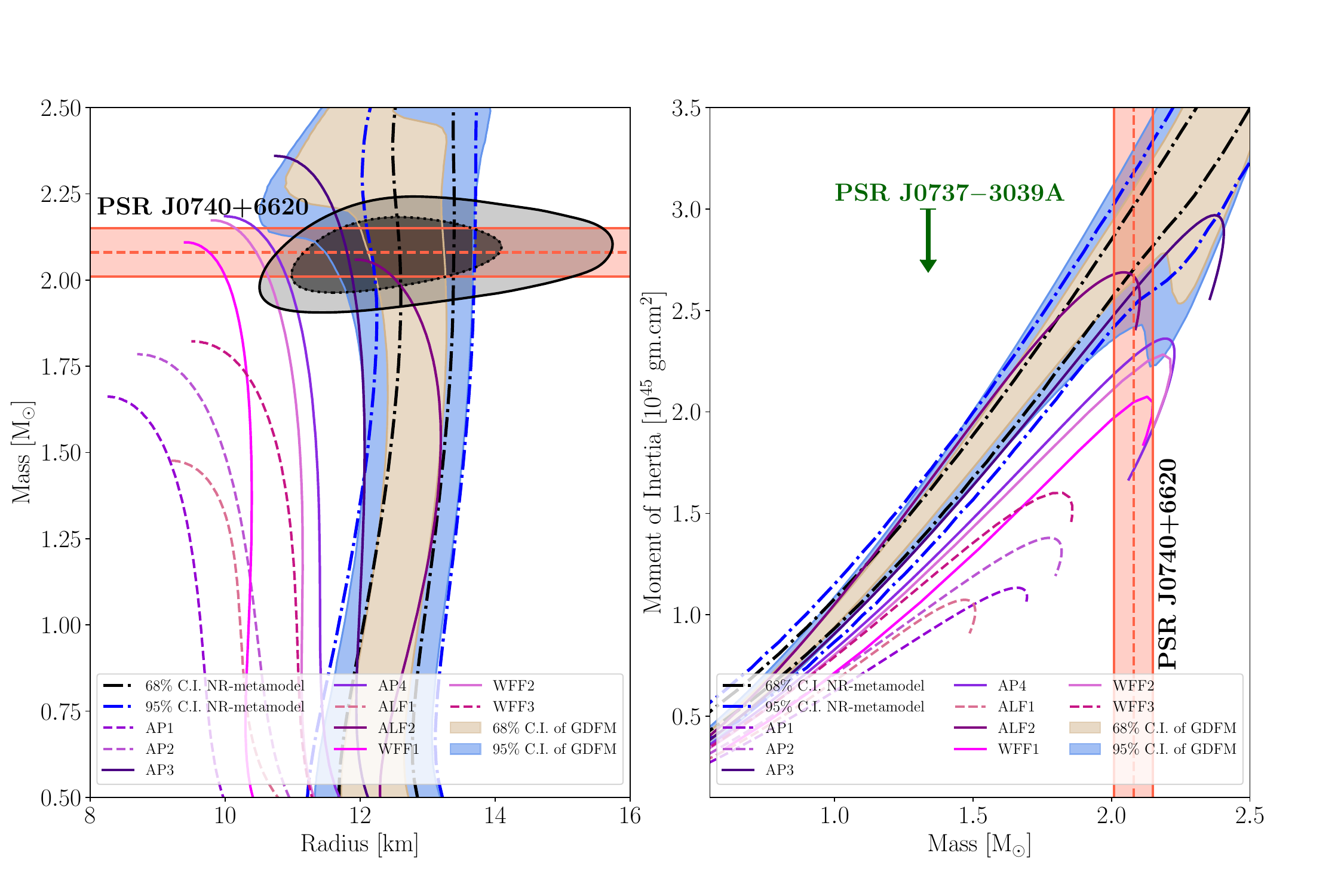}
\caption{\ac{NS}'s $M-R$ (left) and $M-I$ (right) relations for various nuclear \acp{EoS}. Solid lines denote models from \citet{LattimerPrakash2001} with maximum masses exceeding 2.08(7)\,$M_\odot$ as measured for PSR~J0740+6620~\citep{Fonseca:2021wxt}; dashed lines correspond to those that do not. Brown and blue shaded regions show 68\% and 95\% credible intervals from Bayesian inference using relativistic meta-models~\citep{Char:2023fue}, while dash-dotted contours mark equivalent non-relativistic analyses~\citep{montefusco_2025}. The grey region in the left panel indicates the 95\% NICER confidence bounds for J0740+6620~\citep{Salmi24a}, with the inner dashed line showing the 68\% interval. The green arrow marks the 90\% upper limit on $I$ for PSR~J0737$-$3039A~\citep{Kramer:2021jcw}.}

\label{fig:EOS_mass-radius}
\end{figure*}

In general, the radio timing observations summarised in Figure~\ref{fig:NS_masses} now confirm that some \acp{NS} have masses $\gtrsim 2 M_\odot$. Presently, the most massive \ac{NS} determined via pulsar timing is PSR~J0740$+$6620 with a mass of $2.08 \pm 0.07$\,M$_{\odot}$~\citep{Fonseca:2021wxt}. This measurement has had a major impact on the study of ultra-dense matter as any \ac{EoS} that is incapable of sustaining the most massive \ac{NS} observed can be excluded. Consequently, PSR~J0740$+$6620's mass measurement has ruled out the softest \acp{EoS} like those shown by dashed lines in the left panel of Figure~\ref{fig:EOS_mass-radius}. The parameter space for viable \acp{EoS} can be further refined by combining mass estimates with radius or \ac{MoI} measurements. As the former is inaccessible through radio timing, and the latter difficult to extract (see Section~\ref{sec:MoI} for details), complementary constraints on radii and tidal deformabilities from observations in the X-rays and from \ac{GW} mergers, respectively, are invaluable despite their lower precision. As discussed in more detail in Section~\ref{sec:synergies}, the synergy between X-ray and radio observations is particularly noteworthy in this regard as highlighted by the black contours in the left panel of Figure~\ref{fig:EOS_mass-radius} showing PSR~J0740$+$6620's combined mass-radius measurements with NICER~\citep{Salmi24a}.

In light of future SKAO's observations, we also highlight that, more recently, the operation of MeerKAT has enabled a large increase in the number and precision of radio timing measurements of binary pulsar masses in the Southern hemisphere~\citep{Serylak2022,Berthereau:2023aod,Shamohammadi2023,Geyer2023,Gautam2024,Bernadich2024,Padmanabh2024,Chisabi:2025ohh}. Currently, the most massive pulsar detected in the Southern part of the sky is PSR~J1614$-$2230 with a mass of $1.97 \pm 0.04$\,M$_{\odot}$~\citep{Demorest2010}. Once SKAO begins taking data, the impact of these measurements will continue to grow as the timing baselines and corresponding precision increase and new pulsars will be discovered~\citep{Keane01.2026.SKA}.

Additionally, \ac{NS} mass measurements can be obtained from so-called `spider' systems shown in dark blue on the top left of Figure~\ref{fig:NS_masses}. Spiders are binary systems comprising a millisecond pulsar and a low-mass, non-degenerate companion star, typically with orbital periods of less than a day~\citep{Dodge:2024nvl}. Based on the mass of the companion, these spiders are categorised as `black widows' with extremely low mass companions ($M_c < 0.05 M_\odot$) and `redbacks' with higher companion mass ($M_c \gtrsim 0.1 M_\odot$)~\citep{Roberts:2013}. Mass measurements in spiders either rely on the observation of eclipses (primarily in the gamma-rays) or a combination of radio timing with optical spectroscopy. Corresponding mass determinations in the latter scenario are, however, often overestimated~\citep{Clark:2023owb} because they rely on very difficult estimates of the orbital inclinations from multi-wavelength light curves of the irradiated companions, which are heavily dependent on models of the heat distribution on their surfaces~\citep{Voisin2020}.

Despite these uncertainties, Figure~\ref{fig:NS_masses} shows that spider pulsar systems host some of the heaviest \acp{NS}~\citep{Linares:2019aua}, suggesting that even more massive ones may soon be discovered using SKAO's radio timing. Such high \ac{NS} masses remain unconfirmed in \ac{GW} detections (light purple in Figure~\ref{fig:NS_masses}), where precision limits current measurements but will improve with next-generation GW detectors~\citep{Abac:2025saz}. Several compact objects discovered through \ac{GW} mergers (e.g., GW190814; \citealt{Abbott:2020ApJ}) and pulsar timing (e.g., the companion of PSR~J0514$-$4002E; \citealt{Barr:2024wwl}) have masses in the `lower mass-gap' region between $\sim2$–$3,M_{\odot}$, where their nature—\ac{NS} or black hole—remains uncertain. Refined timing and mass measurements from upcoming SKAO's pulsar surveys (Section~\ref{sec:large_surveys}) will provide 
crucial information on the maximum predicted \ac{NS} gravitational mass,
 clarify the status of these mass-gap objects, and tighten limits on \ac{EoS} models, including those predicting deconfinement phase transitions or twin stars~\citep[e.g.,][]{Glendenning:1998ag, Blaschke:2019tbh}.


Beyond maximum-mass constraints, dense-matter properties are also informed by the lightest \acp{NS}. The minimum \ac{NS} mass influences core-collapse formation~\citep{Yasin:2018ckc, Janka2023, Muller2025}, binary evolution~\citep{Tauris2019, You2025}, and the viability of various \acp{EoS}, especially non-nucleonic ones. The lightest confirmed \ac{NS}, the companion of PSR~J0453$+$1559 ($1.174 \pm 0.004,M_{\odot}$; \citealt{Martinez2015}), still permits a broad range of \ac{EoS} families. Candidates below $1\, M_{\odot}$, such as the central compact object in HESS~J1731$-$347~\citep{Klochkov2015, Doroshenko2022}, could challenge current models, though uncertainties in distance and atmospheric models allow for higher inferred masses~\citep{Alford2023}. Future SKAO's timing observations will be crucial to robustly determining the full \ac{NS} mass distribution and validating such low-mass candidates~\citep{Brodie2023, Sagun2023}.

Finally, we note that while the mass measurement techniques outlined above (apart from model-dependent spectral X-ray fitting) only allow constraint of \ac{NS} masses that are located in binary systems, pulsar spin-up glitches are potential (albeit also model-dependent) tools to infer masses of isolated, slowly rotating \acp{NS}~\citep[see, e.g.,][and Section~\ref{sec:glitches}]{Ho_2015_science, Pizzochero2017}.


\subsection{Moment of inertia measurements}
\label{sec:MoI}

Above, we illustrated how the TOV equations define a unique relationship between two macroscopic stellar quantities---mass and radius---for a given \ac{EoS}. Astrophysical observations, such as those from PSR~J0740$+$6620~\citep{Fonseca:2021wxt}, thus constrain a region in the mass–radius diagram, thereby placing limits on viable \ac{EoS} models. However, while pulsar timing provides accurate information about spin frequencies and masses, other bulk parameters such as the radius and the \ac{MoI} are difficult to extract and lower precision measurements of the radius using X-ray data~\citep[e.g.,][]{Riley:2019yda, Miller:2019cac, Salmi24a, Choudhury:2024xbk} are generally used to constrain the \ac{EoS} further. Similarly, the right-hand panel of Figure~\ref{fig:EOS_mass-radius} illustrates the relationship between mass and \ac{MoI}~\citep{Hartle_1968}. In analogy with the $M$–$R$ plane, one can envisage observational constraints on the $M$-$I$ plane, which would offer independent means of probing the \ac{EoS}~\citep{Lattimer2005}.

A general approach to estimating $I$ involves gamma-ray observations of pulsars. Because a significant portion of the pulsar’s spin-down power, given by $\dot E = 4\pi^2 I \dot P/P^3$, is emitted as gamma rays, measured luminosities can give insights into the \ac{NS} \ac{MoI}. However, as the gamma-ray luminosity represents only a fraction of the total spin-down energy, this method yields a lower bound on $I$, typically $> 10^{45}\,$g$\,$cm$^2$. Nonetheless, such constraints are weak due to uncertainties in the distance to the pulsar and the beaming geometry, which make precise \ac{MoI} estimates from gamma-ray data difficult.

It is, however, promising that constraints in the $M$-$I$ plane can (albeit difficult), in principle, be derived solely from high-precision radio timing, one of the key advantages of upcoming SKA observations. In particular, for some systems the timing precision is such that next-to-leading order (NLO) effects (i.e., those beyond standard post-Keplerian effects) become detectable. The foremost example of this is the `double pulsar' PSR~J0737$-$3039A/B. In this system, NLO effects can be detected in the Shapiro delay, aberration, and rate of advance of the orbit's periastron, $\dot{\omega}$~\citep{Kramer:2021jcw}. The latter is the most relevant to us: One of the two measurable NLO contributions to $\dot{\omega}$ is due to relativistic spin-orbit coupling causing the orbital plane to precess about the total angular momentum vector, resulting in $\dot{\omega}_{\rm LT}$. This effect, also known as Lense-Thirring precession, depends on the angular momentum of pulsar A in this system. As the spin of pulsar A is extremely well known---and aligned with the orbital angular momentum~\citep{Ferdman2013}---we can constrain $I$ directly from $\dot{\omega}_{\rm LT}$. Current measurements using $16\,$yr of double pulsar timing data have resulted in an upper limit of $3 \times 10^{45}\, \rm g \, cm^{2}$ at 95\% confidence level on the \ac{MoI} of PSR~J0737$-$3039A~\citep{Kramer:2021jcw}, which will be dramatically improved upon by SKA-Mid observations~\citep[][see also below]{Hu:2020MNRAS}.

Yet improvements in timing precision alone do not address the major roadblock to obtaining accurate $I$ measurements with the SKAO's observations. Systematic uncertainties in the inferred post-Keplerian parameters, namely the orbital period derivative, $\dot{P}_{b}$, from incomplete modelling of acceleration in the Galactic potential can limit the ultimate precision with which the effects of Lense-Thirring precession can be measured. This is presently the major factor limiting \ac{MoI} measurements with the highly-relativistic double \ac{NS} PSR~J1757$-$1854, for which the effects of Lense-Thirring precession is expected to be even stronger than the double pulsar~\citep{Cameron:2017ody, Cameron:2023pfr}. Better models of the Galactic potential based on ongoing analyses of data produced by large-scale astrometry missions such as Gaia~\citep{Sanderson:2016ApJ} may further reduce or even eliminate these uncertainties over the coming decade, setting the stage for major improvements on \ac{MoI} constraints in the era of SKAO.
Moreover, as discussed in detail in Section~\ref{sec:large_surveys}, SKAO will inevitably find more double \ac{NS} systems, and potentially even pulsar black-hole binaries~\citep{Keane01.2026.SKA, Levin01.2026.SKA}, some of which are likely to be more compact than presently known pulsar binaries. Such discoveries would open up new, independent avenues of measuring the \ac{NS} \ac{MoI}. 

For completeness, we also highlight that radio timing of pulsars in binaries is not the only way to constrain $I$. As outlined in Section~\ref{sec:glitches}, observing pulsar glitches provides information on fractional moments of inertia of \acp{NS}, i.e., the fractional contributions of the internal superfluid moments of inertia relative to the total stellar \ac{MoI}~\citep{Graber2018ApJ, Basu:2018fbt, Montoli:2021act, Antonopoulou:2022rpq, Antonelli:2023vpd}. Although these estimates are strongly model dependent, they provide a unique view into the \ac{NS} interior otherwise hidden from view, which are especially powerful when combined with the global constraints outlined above.


\subsection{Maximum spin frequency}
\label{sec:max_spin}

The theoretical maximum spin frequency of a \ac{NS} is equal to the Keplerian frequency, $f_K$, at the stellar surface, i.e., the maximum spin frequency a star can support before mass shedding. To obtain the Keplerian frequency for a given \ac{EoS}, we can perform full \ac{GR} calculations from a general Kerr-like metric. Such efforts have been made by~\citet{Komatsu:1989zz,Friedman1992,Cook:1993qr,Stergioulas:1994ea}.

\begin{wrapfigure}{r}{0.45\textwidth} 
    \vspace{-15pt}
    \includegraphics[width=0.44\textwidth]{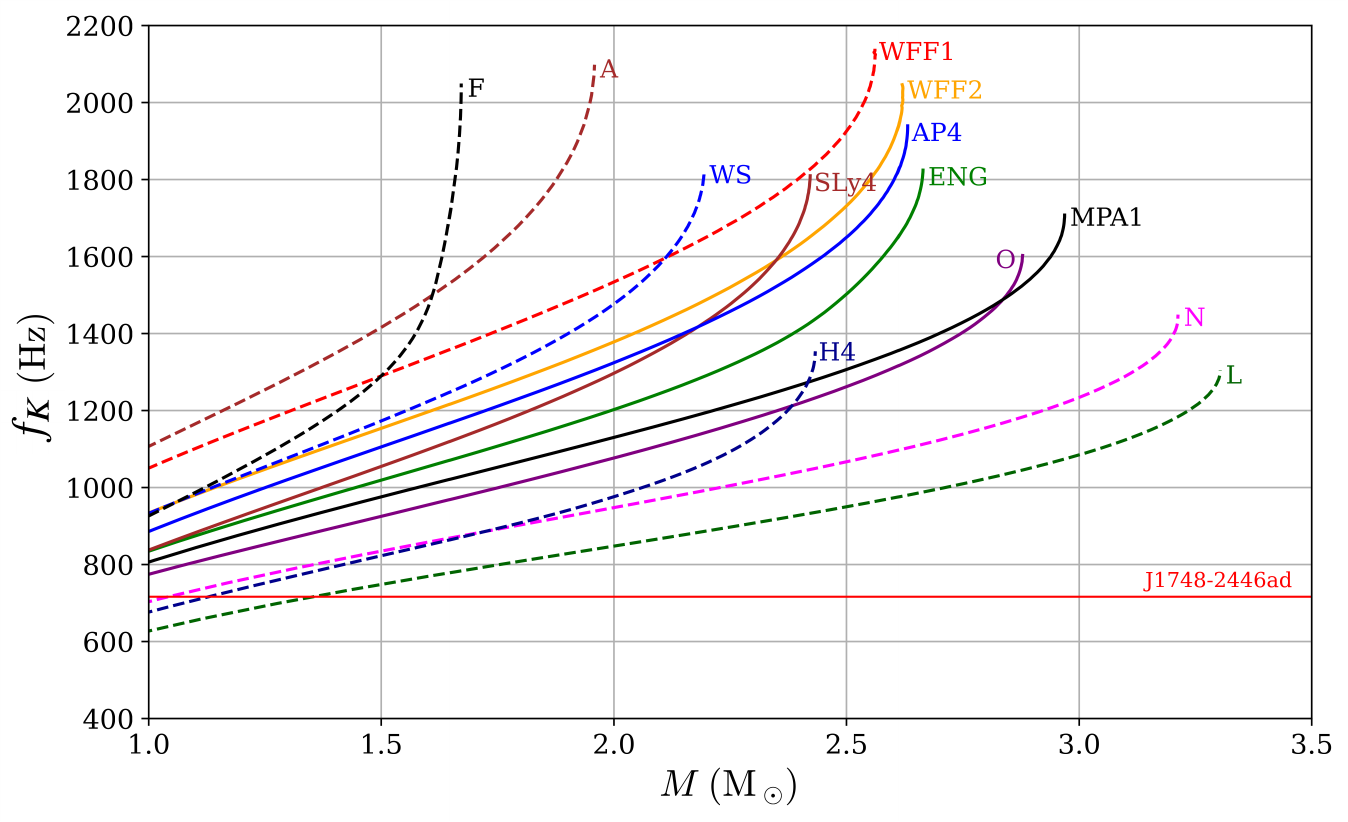}
\vspace{-5pt}
\caption{Maximum spin frequency of \acp{NS} as a function of mass for different \acp{EoS}, computed using the Rotating Neutron Star (RNS) code (\url{https://github.com/cgca/rns}). \ac{EoS} labels follow \cite{LattimerPrakash2001} and the RNS repository~\citep{Stergioulas1996}. Solid curves show maximally rotating configurations consistent with the mass of PSR~J0740+6620~\citep{Fonseca:2021wxt} and multi-messenger constraints~\citep{Dietrich2020}, while dashed curves indicate disfavoured \acp{EoS}. The red line marks the spin frequency of the fastest known pulsar, PSR~J1748$-$2446ad~\citep{Hessels2006}. Higher spin measurements could further constrain the \ac{EoS}. Figure by Norbert Wex.
}
\vspace{-20pt}
\label{fig:mass_spin}
\end{wrapfigure}

The maximum spin frequency of a \ac{NS} is often expressed as $f_K = C\, (M/M_\odot)^{\gamma_1} (R/10\,{\rm km})^{\gamma_2}$, with $\gamma_1=0.5$ and $\gamma_2=-1.5$. Reported values of $C$ vary, e.g. $C\sim 1225$~\citep{Haensel:1989mvc}, $1209$~\citep{Friedman1992}, $1045$~\citep{Lattimer:2004pg}, and $1080$~\citep{Haensel2009}. \citet{Gartlein:2024cbj} showed that reproducing hybrid stars requires $C$ to depend on the deconfinement phase transition rather than remain constant. Since $C$, $\gamma_1$, and $\gamma_2$ are often chosen in an \ac{EoS}-independent way, this oversimplifies reality, as different \acp{EoS} yield distinct compactness values and thus different mass-shedding frequencies. As illustrated in Figure~\ref{fig:mass_spin}, softer \acp{EoS} (more compact stars) reach higher spin limits, while stiffer ones correspond to lower Kepler frequencies (for details see~\citealt{Basu:2024rio}). Similar results using pseudo-Newtonian potential~\citep{Bagchi:2008su} show reduced $f_K$ values for both neutron and quark matter compared to the one derived from the \ac{EoS}-independent analytical expression (given above), possibly hinting towards the scarcity of sub-millisecond pulsars.

To date, the observed spin frequencies of millisecond pulsars lie well below the maxima allowed by realistic \acp{EoS} for the full range of \ac{NS} masses. However, it is important to highlight that the largest spin frequency currently measured for a radio pulsar, $716\,$Hz for PSR~J1748$-$2446ad located in the globular cluster Terzan 5~\citep{Hessels2006}, may already exclude several \ac{EoS} models if its mass were known to be small. This statement is supported by the $M$-$f_K$ relation shown in Figure~\ref{fig:mass_spin}. It is evident that the stiffest \ac{EoS} (green dashed curve labelled L) cannot support a \ac{NS} of mass $\sim 1.3 \, M_\odot$, spinning maximally at a frequency of $716\,$Hz. Thus, if the mass of PSR~J1748$-$2446ad was measured to be smaller than this limit, this specific \ac{EoS} would be ruled out.

Despite being located in a binary system, the prospects of a mass measurement for PSR~J1748$-$2446ad are slim, because it is located in a spider system where relativistic effects are undetectable given the orbital characteristics of the system \citep{Hessels2006}. Complementary estimates at other wavelengths are also unlikely given the current non-detection of the system at optical or near-infrared wavelengths. However, as discussed below in detail, it is possible that future SKAO's pulsar searches~\citep{Keane01.2026.SKA} will find some equally fast or faster-spinning pulsars in systems where mass measurements via radio timing are possible. In such cases, the \ac{EoS} could be significantly constrained even for spin periods below $1000\,$Hz. Thus, as Figure~\ref{fig:mass_spin} shows, the combination of a fast spin with a \ac{NS} mass measurement is a more powerful probe of the \ac{EoS} than a fast spin period alone.

While measuring \ac{NS} spin frequencies does not require a binary, the companion is essential for determining mass. This raises the question: do certain binaries host both massive and rapidly spinning \acp{NS}? Spider pulsar systems, known to contain some of the heaviest \acp{NS}~\citep{Linares:2019aua} as discussed in the Section~\ref{sec:mass}, and shown in the Figure~\ref{fig:NS_masses}. Combining the criteria discussed in the Sec.~\ref{sec:mass} for selecting spider systems with the condition of spin periods $<16\,$ms for identifying millisecond pulsars~\citep{Halder:2023rfu} from the ATNF Pulsar Catalogue\footnote{\href{https://www.atnf.csiro.au/research/pulsar/psrcat/}{https://www.atnf.csiro.au/research/pulsar/psrcat/}}~\citep{manchester:2005atnf}, we find their median spin frequency of spider pulsars is $\sim1.3$ times higher than that of non-spider millisecond pulsars. This suggests spider systems are more likely to host massive, fast-spinning \acp{NS}. Although their mass estimates depend on uncertain optical light-curve modelling~\citep{Dodge:2024nvl}, follow-up radio timing especially with the SKAO will be key to refining orbital parameters and mass determinations.


Most \acp{NS} likely spin well below their Keplerian limit, as several mechanisms halt accretion-driven spin-up in LMXBs—the progenitors of millisecond pulsars. The LMXB spin distribution is bimodal, with a broad slow group and a narrow fast group cutting off near 700 Hz~\citep{Patruno17}, unlike the unimodal distribution of millisecond radio pulsars. Moreover, spin frequencies of millisecond pulsars cut off well below theoretical Keplerian frequencies based on the calculations using the low-density \ac{EoS} and causality~\citep{HaskellZdunik18} at high densities, implying requirement of additional spin-down torques, with \ac{GW} emission being a leading candidate~\citep{Gittins19}. Transitional systems showing both accretion-powered X-rays and radio emission in quiescence~\citep{Papitto2022, HaskellPatruno17} may provide key evidence for such \ac{GW} mechanisms.

In fact, the lack of radio pulsars in the lower left quadrant of the $P$-$\dot{P}$ plane (rapidly rotating with low spin-down rates) has been shown to be well modelled in terms of \ac{GW} torques, due to a small residual ellipticity possibly held in place by a buried superconducting magnetic field~\citep{Woan18}. A more detailed understanding of the spin distribution of the millisecond pulsar population---a key science goal for the SKAO~\citep{Keane01.2026.SKA}---will, hence, allow us to understand not only the evolution of magnetic fields and the electromagnetic emission of \acp{NS} in the era of SKAO's discoveries, but also the evolution of their quadrupoles and \ac{GW} emission, informing \ac{GW} searches with current and future ground based interferometers~\citep{Haskell15}.


\subsection{Pulsar glitches} 
\label{sec:glitches}

\begin{figure*}[t!]
\centering
\includegraphics[width=0.95\linewidth]{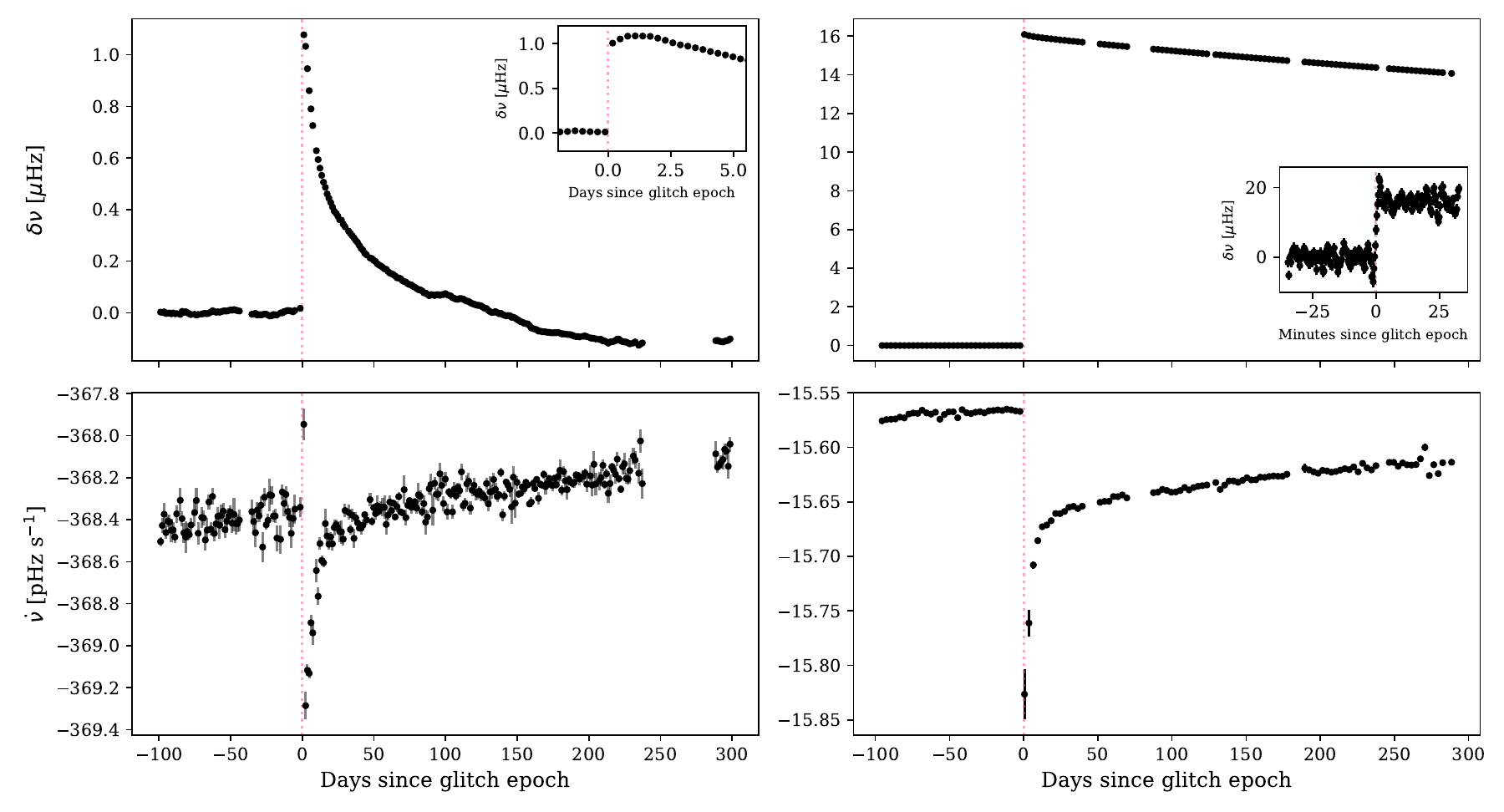}
\caption{Evolution of the spin-frequency (upper row) and the spin-down rate (lower row) of PSR~B0531$+$21 (the Crab pulsar, left column) and PSR~B0833$-$45 (the Vela pulsar, right column) close in time to their respective large glitches in 2019 and 2016. The values were computed using a striding boxcar method (e.g., \citealt{Shaw:2018MNRAS}) in which a fit for both parameters was applied to consecutively overlapping groups of times-of-arrival. The insets show the more densely sampled frequency evolution around the glitch epoch (vertical dotted line).}
\label{fig:crabvelaf0f1}
\end{figure*}

Glitches are sudden increases in a pulsar's rotational spin frequency, which are associated with the rapid angular momentum transfer from the superfluid interior to the observed component of a \ac{NS}~\citep{Antonopoulou:2022rpq,Zhou:2022}. As discussed in Section~\ref{sec:unknowns_SFphysics}, this transfer occurs when many previously pinned vortices unpin and migrate outwards, spinning down the superfluid while accelerating the \ac{NS} crust~\citep{haskell2015IJMPD,Antonopoulou:2022rpq} leading to an observed step $\Delta \nu>0$. In most events, the spin-down rate $|\dot{\nu}|$ also increases abruptly ($\Delta\dot{\nu}<0$), as internal components decouple, and the relaxation of the crust-interior system is reflected in the observed rotation for a duration referred to as `post-glitch recovery'. These sudden spin-ups provide access to dense-matter physics, as glitch characteristics can reveal the \ac{NS} structure (e.g., the amplitudes for the $\nu$ and $\dot{\nu}$ changes depend, among others, on the relative \ac{MoI} between the various components), internal transport properties and microphysical processes (which, for example, dictate the relaxation timescales). See Section~\ref{sec:unknowns_SFphysics} for further details.

Glitches span amplitudes of $10^{-10} \lesssim\Delta\nu \lesssim 10^{-5}\,$Hz\footnote{See the \ac{JBO} Glitch Catalogue~\citep[][\href{https://www.jb.man.ac.uk/pulsar/glitches.html}{https://www.jb.man.ac.uk/pulsar/glitches.html}]{Basu:2021pyd}
}. and occur on timescales from seconds for the spin-up to years for recovery. Figure~\ref{fig:crabvelaf0f1} illustrates this range with two glitches from the Crab and Vela pulsars, showing differences in the spin and the spin-down rate evolution and the recovery behaviour. Crab’s spindown rate relaxes quasi-exponentially within a year, whereas Vela exhibits long-term post-glitch changes persisting until the next event. Despite such diversity, population studies reveal consistent trends in glitch activity as the number of detected events grows~\citep{espinoza2011,fuentes2017,lower:2021mnras,Basu:2021pyd}. High-quality, high-cadence observations are thus essential to resolve underlying physics. With its sensitivity and monitoring capability, the SKAO is ideally positioned to detect more glitches and use them as precision probes of fundamental \ac{NS} properties.

Most glitch rises are unresolved by current observations. In three Vela glitches, the spin-up occurred within tens of seconds (e.g., Figure~\ref{fig:crabvelaf0f1}, top right) and was immediately followed by a fast (minute-long) internal response likely originating from the star's core~\citep{Palfreyman2018,Ashton:2019Nat,Graber2018ApJ,pizzochero2020}. However, for six of the largest glitches of the Crab pulsar, an initial unresolved spin-up was observed to be followed by an extended phase lasting many hours (e.g., Figure~\ref{fig:crabvelaf0f1}, top left)~\citep{lyne:1992nat,wong:2001apj,ge:2020apj,Basu:2019iam, Shaw:2018MNRAS,Shaw:2021MNRAS}. This rare feature of a slow rise is also seen once in each of the radio-quiet magnetars 1E~2259$+$568~\citep{woods:2004apj} and SGR~J1935$+$2154~\citep{ge:2022ar}. The detection of glitch rises in the Crab and Vela pulsars is, to a large extent, possible due to the observing resources afforded to these NSs~\citep{lyne:2015mnras,Dodson2007}. Such observations of the spin-up and subsequent shortest relaxation features (seen in Vela) inform us of the dynamics of superfluid vortices inside \acp{NS}~\citep{AntonelliHaskell2020,sourie:2020mnras} and the role of the star's core in glitches~\citep{Graber2018ApJ,haskell:2018mnras}, while also providing critical constraints on crustal entrainment and the \ac{MoI} of superfluid components~\citep{andersson2012PhRvL,chamel2013PhRvL,Montoli:2018fqz,Montoli:2020A&A}.

\begin{wrapfigure}{r}{0.5\textwidth} 
    \centering
    \vspace{-15pt}
    \includegraphics[width=0.5\textwidth]{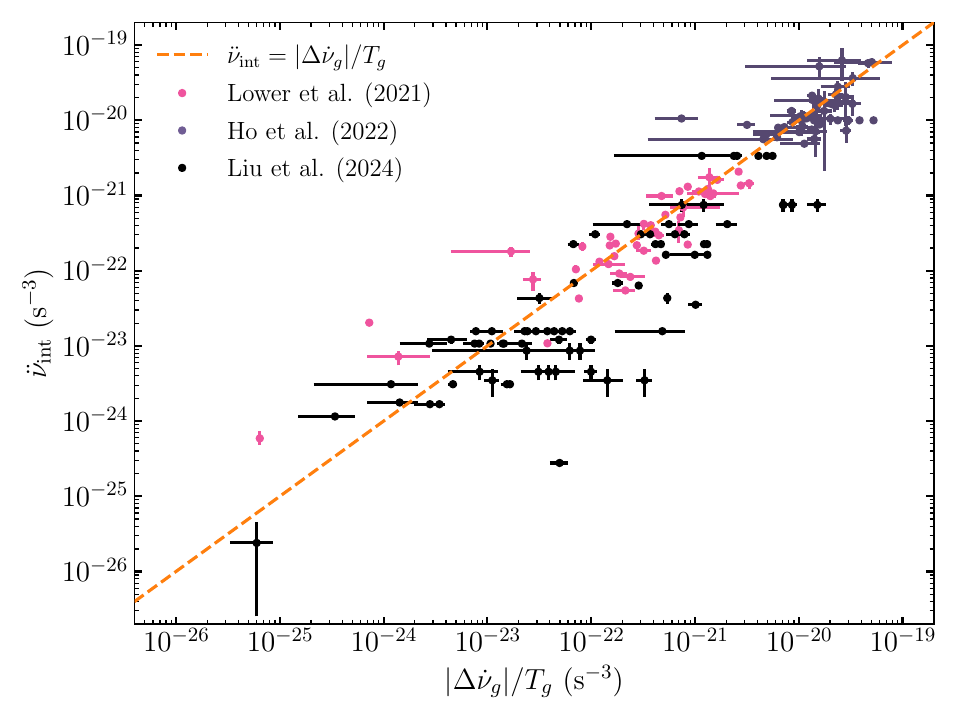}
    \caption{Observed relationship between the second spin-frequency derivative measured between glitches ($\ddot{\nu}_{\rm int}$) and the glitch-induced step-change in spin-down rate normalised by the wait time to the next glitch ($|\dot{\nu}_{g}|/T_{g}$). Points in pink are those presented in \citet{lower:2021mnras}, purple ones are from \citet{Ho:2022kga} for PSR~J0537$-$6910, and in black are the values from \citet{Liu2024} (note, the $\ddot{\nu}$ values from this work are the average across multiple glitches). The dashed orange line indicates the one-to-one relationship which would suggest a complete recovery of $\dot{\nu}$ at the time of the next glitch.}
\label{fig:glt_f2}
\end{wrapfigure}

Regular observations will enable detailed studies of long-term post-glitch relaxations and increase the number of detected glitches. The former can probe weakly coupled superfluid components and reveal internal conditions underlying variability in glitch behaviour~\citep{celora:2020mnras,haskell20}, informing models of rotational evolution in young pulsars. The latter is essential for statistical studies: glitch size and waiting-time distributions vary among pulsars~\citep{melatos:2008apj, fuentes2017}, reflecting differences in angular momentum reservoirs and microphysical properties such as pinning forces, entrainment, and superfluid gaps~\citep{andersson2012PhRvL, Ho_2015_science}. Some pulsars, such as the Crab, show stochastic behaviour, while others, like Vela, exhibit characteristic sizes and intervals. In PSR~J0537$-$6910, the next glitch can be predicted from the previous one, pointing to a threshold-dominated trigger~\citep{Antonopoulou:2017hwa, Melatos2018ApJ}. Distributions and correlations within and across pulsars—e.g. between glitch activity and inferred age—provide key constraints on glitch mechanisms~\citep{Antonopoulou:2022rpq, Antonelli:2023vpd, lower:2021mnras, Liu2024}. Current analyses are sample-limited, but the sensitivity of SKAO’s telescopes and the cadence of observations will enable much larger, more statistically powerful studies (see Section~\ref{sec:glitchobs}).

To fully leverage SKA's capabilities and maximise the scientific gains, a coordinated observational strategy is essential (detailed in Section~\ref{sec:SKA_expect}), with flexibility to change which sources take priority for high-cadence observations, e.g., when a glitch happens or is expected for a particular pulsar.


\subsection{Free precession}
\label{sec:precession}

Free precession offers another point of connection between SKAO's pulsar observations and the \ac{NS} structure. If modelled as a simple biaxial rigid body, a misalignment between the star's (fixed) angular momentum vector and its symmetry axis results in a periodic modulation of the rate and latitude at which the magnetic axis sweeps around the angular momentum vector (see, e.g.,~\citealt{Jones2001,gao_etal_23}). If the \ac{MoI} tensor is diagonal and has components $(I_x, I_x, I_z)$, then, for small angle free precession of a nearly spherical star of spin period $P$,  a long-period modulation $P_{\rm mod}$ is induced in all aspects of the pulsar emission, i.e., in the pulse timing, beam shape, and polarisation, with 
\begin{equation}
    \frac{P}{P_{\rm mod}} \approx \frac{|I_z - I_x|}{I_z} \equiv \epsilon .
\end{equation}
The precession modulates the rate at which the magnetic axis swings past the observer's line of sight, i.e., modulates the observed spin frequency. It also modulates the angle between the star's (fixed) angular momentum vector and the pulsar beam, resulting in variations in the angle through which the 
observer's line of sight cuts the pulsar beam.  This will produce variations in the beam shape and the observed sweep of polarisation with rotational phase.  Also, providing (as is likely) the spin-down torque is sensitive to this latitudinal angle, there will be variation in the spin-down torque, amplifying the variations in spin frequency~\citep{cordes_93,Jones2001}. 

The modelling of \ac{NS} crusts suggests that the ellipticity is limited to values $\epsilon \lesssim 10^{-6}$~\citep{Jones2001, Johnson2013, Gittins21a, Gittins21b}, so that a typical $P \sim 1$\, second pulsar would display modulations on the timescale of months. However, the connection between observation and theory is complicated by the realisation that if a pinned superfluid component exists, as required to explain pulsar glitches (see Sections~\ref{sec:unknowns_SFphysics} and~\ref{sec:glitches}), then the modulation period is drastically reduced  $P/P_{\rm mod} \approx I_{\rm SF}/I_z$,~\citep{Shaham1977},
where $I_{\rm SF}$ is the \ac{MoI} of the pinned superfluid component, believed to be of the order a few percent of the total stellar \ac{MoI}. It follows that the observable modulations might appear on timescales as short as tens of seconds for a typical $P\sim 1$\,s pulsar. Furthermore, in this scenario, the precession may itself be damped rather rapidly, due to the dissipative interaction of the neutron superfluid vortices with the star's magnetic field~\citep{link_06}. Clearly, there is a wide range of theoretical possibilities, demanding the analysis of data acquired by SKAO with both high cadence and long duration.

A number of pulsars have already been seen to display quasi-periodic oscillations in their spin-down rates, with free precession sometimes being advanced as the cause, the cleanest example being PSR~B1828$-$11~\citep{Stairs2000, Ashton2016, Ashton2017}. However, subsequent analysis has revealed that in some cases at least, these timing variations are accompanied by sharp changes in emission profile, something not expected for free precession~\citep{Lyne2010, Stairs2019}. It has been argued that this may reflect the pulsar magnetosphere~\citep{Oswald01.2026.SKA} being finely balanced between two states, with the changing geometry of the free precession providing a statistical bias as to which state is preferred at any given precessional phase~\citep{Jones2012}.  However, the origin of the timing variations themselves remains unclear.

Further progress on understanding the origin of these quasi-periodic oscillations will require analysis of contemporaneous spin-down and pulse profile data, ideally for a large set of pulsars. This is particularly important given the recent results of~\citet{lower_etal_25}, who found that such correlated changes may be more common than previously thought. Carrying out this analysis for as large a set of pulsars as possible is crucial, as the statistical correlation (or lack thereof) between the pulsars' spins and modulation periods potentially contains important information on the mechanism at work in producing the periodicity~\citep{Jones2012}.

Furthermore, there are potential connections between free precession and the glitches described in Section~\ref{sec:glitches}. First, as noted above, the storage of pinned vorticity in a neutron superfluid may produce extremely short modulation timescales~\citep{Jones:2016oyh}.  Second, it is possible that non-axisymmetries in pulsar glitches may themselves `kick' the star into free precession~\citep{jones_andersson_02}, with a crust-cracking event suddenly changing the \ac{MoI} tensor, instantaneously producing a misalignment between the (fixed) angular momentum vector and a principal axis.  This opens up the need to record timing data in the time interval \emph{immediately following} a glitch, something which the SKAO should be well placed to achieve.


\section{Potential impact of dark matter and modified gravity models}
\label{sec:DM_modgrav}

Given that little is known about one quarter of the total energy of the Universe, the dark sector might affect pulsar observations. While this field of research is relatively new, studies show that the impact of Beyond the Standard Model physics and gravity beyond \ac{GR} show a degeneracy with \ac{EoS} uncertainties, hindering us from probing the baryonic matter \ac{EoS}~\citep{Yazadjiev:2014cza,Giangrandi:2022wht}.
Due to their extreme compactness, \acp{NS} can accumulate a sizeable amount of \ac{DM} particles from the surrounding galactic medium. This is especially important for the observations of pulsars close to the Galactic centre (see also~\citealt{Abbate2025_SKA_GalCen}), where the \ac{DM} density is significantly higher than far away from the centre, a feature referred to as the \ac{DM} spike~\citep{Ullio:2001fb}. 
If the interaction between the visible and dark sectors other than gravity is very weak, the components do not exist in equilibrium. In this case, the \ac{DM}-admixed NSs are defined by two pressures and two energy densities, one for each of the components. Additionally, the amount of \ac{DM} in each star may vary depending on the \ac{DM} density in the surrounding medium, which results in a two-dimensional plane, i.e., \ac{DM} fraction and particle mass or interaction strength ~\citep{Hippert:2022snq}.

\begin{figure*}[t!]
\centering
\includegraphics[height=0.45\linewidth]{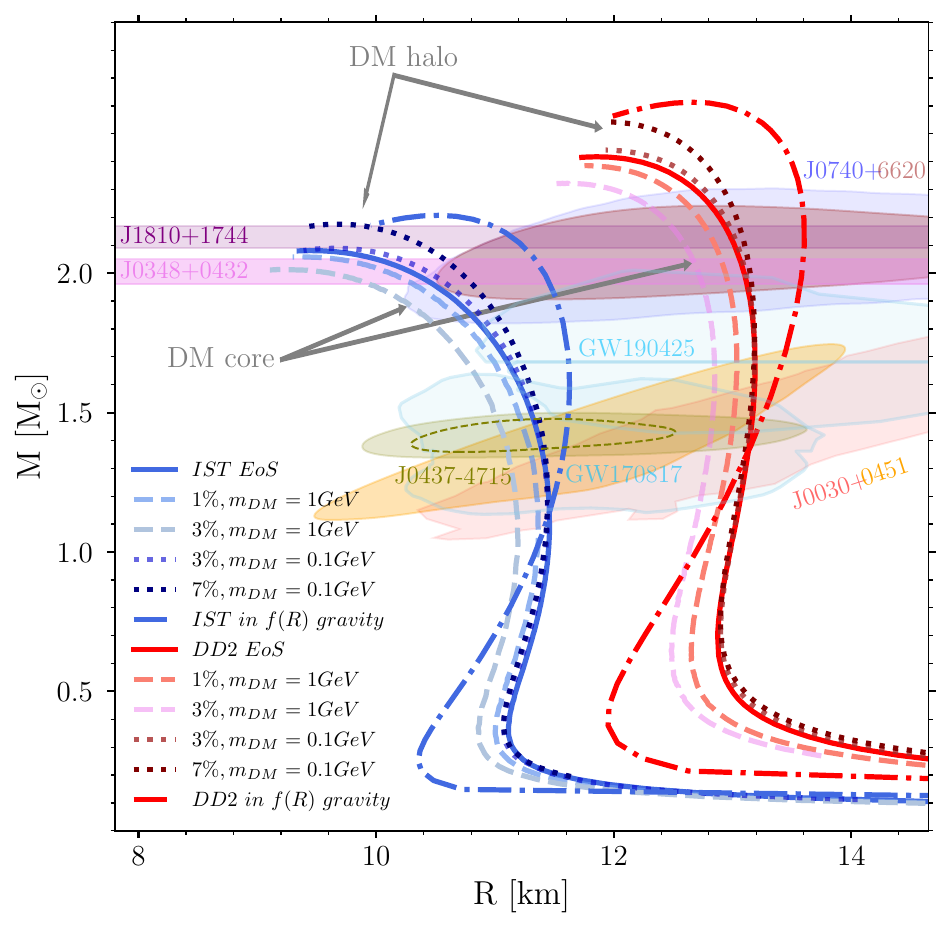}
\includegraphics[height=0.45\linewidth]{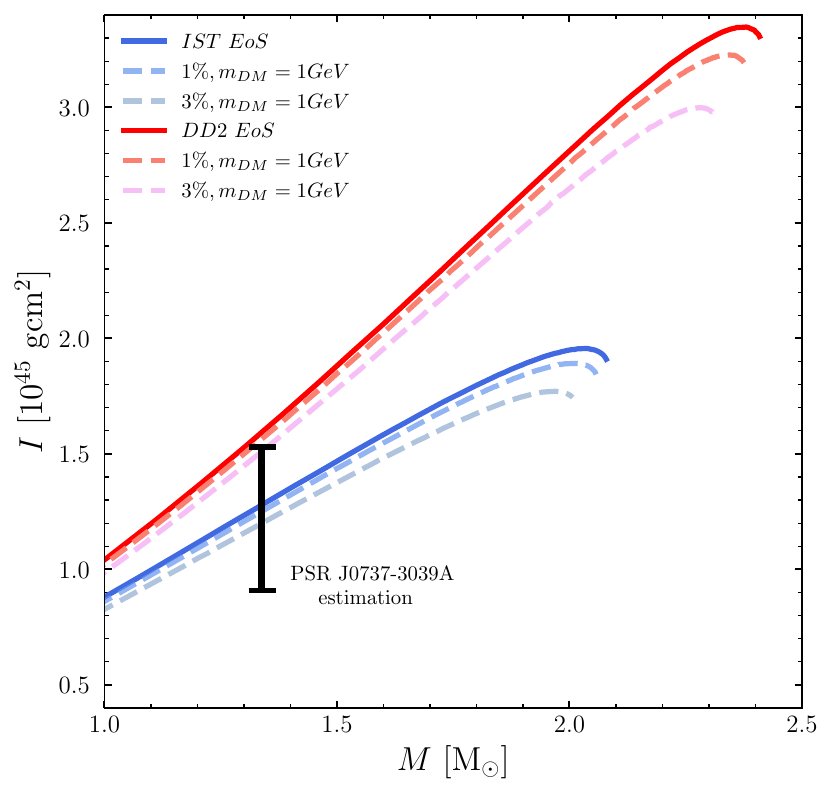}
\caption{Total gravitational mass as a function of the visible baryonic radius (left panel) and the \ac{MoI} as a function of the total gravitational mass (right panel) for \ac{DM}-admixed \ac{NS} is shown for several \ac{DM} fractions, and \ac{DM} particle masses, $m_{\rm DM}$. The results were obtained for asymmetric fermionic \ac{DM} that interacts with the visible sector only through gravity (for more details see~\citealt{Ivanytskyi:2019wxd}). To address the baryonic matter \ac{EoS} uncertainties, we consider the soft IST \ac{EoS}~\citep{Sagun:2020qvc} (solid blue curve) and stiff DD2~\citep{Typel:1999yq} (solid red curve) as two extreme limits~\citep{Cipriani:2025tga}. The dashed and dotted curves depict the core and halo configurations, respectively, considering \ac{GR}. The $M(R)$ dash-dotted blue and red curves were obtained for the $f(R)= R+\alpha R^{2}$ gravity for $\alpha$=100~\citep{Yazadjiev:2014cza, Yazadjiev:2015zia}. The 1$\sigma$ confidence interval constraints from GW170817, GW190425, the NICER measurements of PSR~J0030$+$0451~\citep{Miller:2019cac, Vinciguerra:2023qxq}, PSR~J0740$+$6620~\citep{Dittmann24,Salmi24b}, and PSR~J0437$-$4715~\citep{Choudhury:2024xbk}, as well as mass measurements of heavy radio pulsars (PSR~J1810$+$1744, PSR~J0348$+$0432) are also plotted. The estimated $I$ of PSR~J0737$-$3039A (without accounting for \ac{DM} or modified gravity) with 90\% confidence interval is depicted in black in the right panel~\citep{Landry:2018jyg}.}
\vspace{-15pt}
\label{fig:DM}
\end{figure*} 

The impact on the \ac{NS} properties, such as gravitational mass, radius, \ac{MoI}, matter distribution, etc., depends on the \ac{DM} candidate, its particle mass, and self-interaction strength~\citep{Bramante:2023djs,Cipriani:2025tga}. Very weakly interacting heavy \ac{DM} particles tend to form a dense core inside the baryonic \ac{NS} leading to a smaller radius for a given gravitational mass of the star, which is related to the additional gravitational pull of \ac{DM}~\citep{Leung:2011zz,Ivanytskyi:2019wxd,Giangrandi:2022wht,Barbat:2024yvi}. This effect mimics the softening of the \ac{EoS} caused by the appearance of new heavy degrees of freedom including deconfined quarks~\citep{Biesdorf:2024dor}. On the other hand, light \ac{DM} can form an extended halo around the baryonic \ac{NS} that increases the star's total gravitational mass and resembles the stiffening of the \ac{EoS}~\citep{Nelson:2018xtr,Karkevandi:2021ygv}. A similar effect is obtained for axions and other ultra-light particle clouds surrounding a \ac{NS}~\citep{Noordhuis:2023wid}. The corresponding $M(R)$ relations for different \ac{DM} configurations and the impact of \ac{DM} on $I$ are shown in Figure~\ref{fig:DM}.
In this \ac{DM} analysis, \ac{GR} was assumed. However, it has been shown that a modification of gravity itself can also impact the \ac{NS} properties and, consequently, inferred quantities such as mass and radius~\citep{Shao:2019gjj}. As illustrated in the left panel of Figure~\ref{fig:DM}, the $M(R)$ relations obtained for \ac{GR} and $f(R)=R+\alpha R^2$ gravity~\citep{Starobinsky:2007hu} exhibit a high degree of degeneracy making it difficult to disentangle the effects. Therefore, measuring only the \ac{NS} radius at a given mass would not be sufficient to distinguish between the effects of \ac{DM} or modifications of gravity and the dense matter properties.

Although \acp{NS} provide a compelling testing ground for gravity, nuclear physics, and physics Beyond the Standard Model, the possible degeneracy between the effects of \ac{DM} and/or gravity beyond \ac{GR} and dense matter properties could lead to misleading conclusions when analysing the SKAO's data in isolation. Therefore, the joint efforts of SKAO and other facilities to obtain multi-messenger observations of \acp{NS}, along with advances in experimental and theoretical subatomic physics are pivotal for breaking these possible degeneracies and shedding light on the \ac{NS} internal composition.

\section{Expectations in the SKAO era}
\label{sec:SKA_expect}



In this Section, we discuss the impact of the SKAO on advancing our understanding of ultra-dense matter around and beyond nuclear saturation density.


\subsection{Advances due to SKAO's improved sensitivity}

The sensitivity of the SKAO's array will increase as additional antennas are commissioned, substantially extending its scientific reach. In the AA* and AA4 configurations, SKA-Mid will be approximately three and four times more sensitive than MeerKAT, respectively. SKA-Low will ultimately exceed the sensitivity of LOFAR by nearly an order of magnitude.

The increased sensitivity of the SKAO's arrays will directly reduce uncertainties on pulsar \acsp{ToA}, enabling unprecedented precision in mass measurements for both known and newly discovered binaries. SKA-Mid, in particular, will provide the tightest constraints yet on the maximum \ac{NS} mass and, due to its broad frequency coverage, will allow constraints on pulsar orbital parameters via scintillation measurements.

\begin{wrapfigure}{l}{0.5\textwidth} 
    \centering
    \vspace{-20pt}
\includegraphics[width=0.5\textwidth]{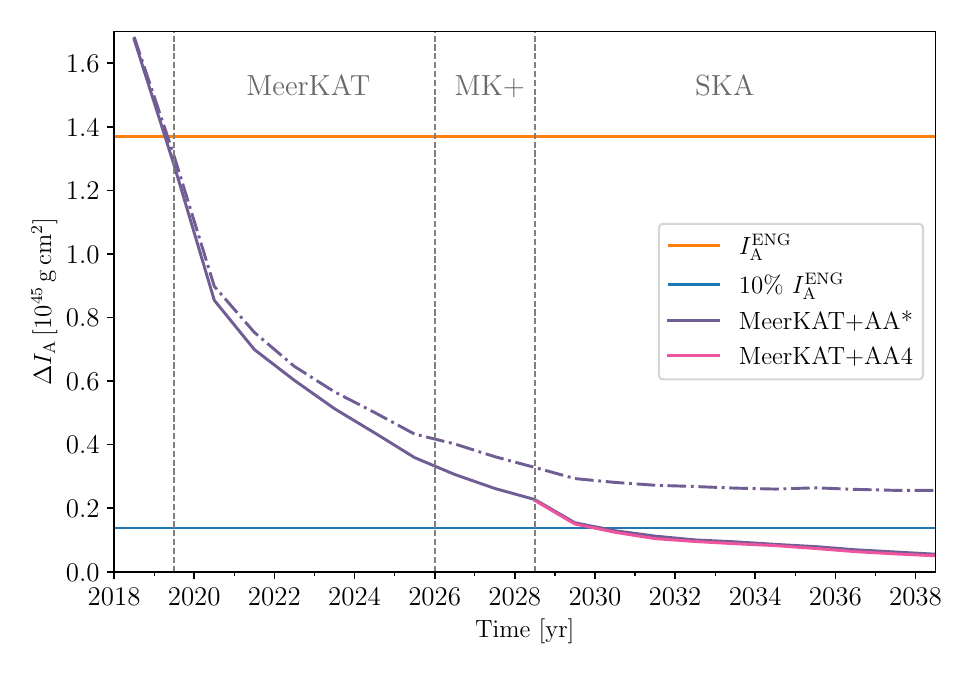}
\vspace{-15pt}
\caption{Simulated uncertainty in the \ac{MoI} of pulsar~A in J0737$-$3039 versus observing time with MeerKAT, MeerKAT+ (\href{https://www.meerkatplus.tel/}{https://www.meerkatplus.tel/}), and SKA. Simulations assume the ENG \ac{EoS}~\citep{LattimerPrakash2001} and $3$\,hr monthly observations with full arrays. The orange line marks the theoretical $I_{\rm A}$, and the blue line an uncertainty of 10\%. The purple dash-dotted line shows the expected uncertainty evolution from recent Galactic measurements~\citep{Gravity2021, Guo2021}, flattening after 2030. Solid lines (purple: AA*, pink: AA4) indicate improved predictions assuming refined Galactic parameters and enhanced timing parallax and proper motion.
}
\vspace{-20pt}
\label{fig:MoI_uncertainty}
\end{wrapfigure}

The increase in sensitivity will similarly set unprecedented constraints on the \ac{NS} \ac{MoI}. The most promising existing candidate is the double pulsar PSR~J0737$-$3039A/B (see right panel of Figure~\ref{fig:EOS_mass-radius}). To estimate corresponding uncertainties with time, we can simulate the expected precision of the three relevant post-Keplerian parameters (the advance of periastron $\dot{\omega}$, the Shapiro parameter $s$, and the rate of change of orbital period $\dot{P}_\mathrm{b}$) with MeerKAT and SKAO's telescopes (see~\citealt{VenkatramanKrishnan01.2026.SKA} for details on these simulations). Subsequently removing the extrinsic kinematic contributions and performing the $\dot{P}_\mathrm{b}$--$\dot{\omega}$--$s$ test using the intrinsic contributions for the former two quantities as outlined in~\citet{Hu:2020MNRAS} leads to uncertainties in the \ac{MoI} for pulsar A of $10-23$\% by 2030 and $4-20$\% by 2038 (after 10 years of timing with the SKA) at 68\% confidence, respectively. These ranges are determined by our uncertain knowledge of the underlying Galactic potential as discussed in Section~\ref{sec:MoI} and highlighted in Figure~\ref{fig:MoI_uncertainty}.

Achieving the required precision to realise new constraints of mass and moment-of-inertia demands timing key pulsars for several hours per month with SKA-Mid to build long timing baselines. Although SKA-Low will detect many new pulsars, sources of particular interest must be followed up with SKA-Mid to mitigate pulse broadening and dispersion variability caused by propagation through the ISM at low frequencies.

\subsection{The role of large surveys}
\label{sec:large_surveys}

Pulsar science in the SKAO era will advance through both increased sensitivity and new survey programmes. The broad frequency coverage will enable the detection of faint, previously unknown sources across diverse range of environments. Although low-frequency observations are favoured due to the steep spectra of pulsars,~\citep{Posselt:2022vrk}, scattering and dispersion in the ISM, especially towards the Galactic Centre~\citep{Rickett1977}, make higher frequency observations essential. SKA-Mid will therefore play a key role in probing the inner Galaxy, while SKA-Low surveys the wider sky~\citep{Gonthier2018, Mishra-Sharma2022, Abbate2025_SKA_GalCen, Keane01.2026.SKA}. Population models predict detections of $\sim 10^4$ slow pulsars, $\sim800$ MSPs, and $\sim 110$ double \ac{NS} systems in AA*, with a $\sim20\% $ increase in AA4~\citep{Keane01.2026.SKA}. The SKAO's pulsar search programme is also expected to expand the sample of young, high spin-down power pulsars, which are prime glitch candidates. Of the $\sim200$ glitching \acp{NS} currently known, about 140 are younger than $10^6$ yr and account for $\sim600$ of $\sim700$ observed glitches~\citep[see][and the Jodrell Bank Glitch Catalogue\footnote{\href{http://www.jb.man.ac.uk/pulsar/glitches.html}{http://www.jb.man.ac.uk/pulsar/glitches.html}}]{Basu:2021pyd}. Simulations suggest that SKAO will detect $\sim 600$ such young \acp{NS} in AA*, rising to $\sim 650$ in AA4, providing a rich new glitch sample.


Expanding the known \ac{NS} population will increase the number of systems suitable for nuclear physics tests. Of the $\sim3{,}800$ known radio pulsars, a small fraction of all that are beamed towards Earth~\citep{Graber:2023jgz}, only $\sim10\%$ are in binaries~\citep[ATNF Pulsar Catalogue v2.6,][]{manchester:2005atnf}. Just a few dozen currently allow mass measurements (Figure~\ref{fig:NS_masses}), and only a handful are massive enough to constrain the \ac{EoS}~\citep{Antoniadis:2013pzd, Fonseca:2021wxt}. A larger sample may reveal additional $>2M_\odot$ systems, providing new constraints. Targeted searches of unidentified Fermi sources~\citep{Camilo:2015caa} and Galactic globular clusters~\citep{Bagchi01.2026.SKA}, which favour millisecond pulsars, will be especially valuable. More broadly, mapping the \ac{NS} mass distribution, including the minimum mass, is critical for constraining exotic \acp{EoS}~\citep{Sagun2023}, understanding \ac{NS} formation~\citep{Yasin:2018ckc, Janka2023}, and modelling compact binary evolution~\citep{You2025} and \ac{GW} signals from mergers~\citep{Harry:2018hke, LIGOScientific:2018cki}.

A double \ac{NS} system analogous to the double pulsar but with an orbital period below one hour could enable a \ac{MoI} measurement with $\sim 10\%$ precision after six years of SKAO's timing, improving to $\sim 1\%$ after a decade~\citep{Hu:2020MNRAS}. Similar measurements may be feasible in pulsar–black hole binaries—an SKAO's key science goal~\citep{VenkatramanKrishnan01.2026.SKA}, where stronger gravitational effects~\citep{Liu:2014uka} should allow comparable precision on shorter baselines~\citep{bagchi2013}. Large-scale surveys will also refine our understanding of pulsar spin distributions~\citep{Levin01.2026.SKA}. Of $\sim 3{,}800$ known pulsars, $\sim 640$ are MSPs with $P<30$ ms, and $\sim 560$ spin faster than $10$ ms~\citep[ATNF Pulsar Catalogue v2.6,][]{manchester:2005atnf}. While this sample does not yet exclude specific \acp{EoS} (Figure~\ref{fig:mass_spin}), discovering a sub-millisecond pulsar with a mass measurement would set new constraints on the ultra-dense matter \ac{EoS} (see Section~\ref{sec:max_spin}).

Predicting the expected number of these new system is challenging, but every single such discovery with SKAO has the potential to transform our knowledge of astrophysical constraints of dense matter.


\subsection{Achievable ToA precision with the SKAO -- A case study}
\label{case_study}

%
%
%

The precision of a \ac{ToA} measurement can be approximated by $\sigma_{\rm ToA} \approx \frac{W}{2 {\rm (S/N)}}$ \citep{Bailes:2018azh} where ${\rm S/N}$ is the signal-to-noise ratio of the folded pulse profile and $W$ is the pulse width. For SKA-Low, ${\rm S/N}$ scales directly with the number of identical elements used in beamforming. For SKA-Mid, subarrays may comprise different proportions of MeerKAT and SKAO's dishes, leading to \citep{Basu2025_SKA_EOS},
\begin{equation} \label{radiometer}
{\rm S/N} = S_{\rm mean} \sqrt{\frac{P-W}{W}} \frac{(M G_M + N G_S)^2 \sqrt{N_p \Delta f T_{\rm obs}}}{(M G_M T_M + N G_S T_S)}.
\end{equation}

Here, $S_{\rm mean}$ is the mean radio flux density of the pulsar at a given frequency $f$, $P$ is the spin period of a pulsar, $N_{\rm p} = 2$ is the number of summed polarisations, $\Delta f$ is the bandwidth, $T_{\rm obs}$ is the total integration time of an observation, $M$ is the number of MeerKAT dishes each with antenna temperature gain $G_M$ and system temperature $T_{M}$, and $N$ is the number of SKA dishes each with antenna temperature gain $G_S$ and system temperature $T_{\rm S}$. 
Equation~\eqref{radiometer} reduces to the standard formula~\citep{lorimerkramerhandbook} when beamforming is achieved through an array in which all elements are of the same type. 
We adopt values of $G_M = 0.042\,$K/Jy, $T_M = 20$ K~\citep{Bailes:2018azh}, $G_S = 0.058\,$K/Jy and $T_S = 13.5$ K~\citep{Pellegrini+2020} and assume that SKA-Mid observations
are carried out at L-Band\footnote{`L-Band' refers to the range of frequencies around $1420\,$MHz. In the case of the SKAO, this is covered by Band 2, which ranges from $950-1760\,$MHz.}. 

In order to estimate the achievable \acs{ToA} precision at each array assembly, we adopt a range of fiducial pulsar parameters \citep{Basu2025_SKA_EOS}. We select a pulse profile width of $2.5\,$ms, a pulse period of $300\,$ms (corresponding to a rotation frequency of $\nu = 3.3$ Hz) and a frequency derivative of  $ \dot{\nu} = -1.8 \times 10^{-12}\,$Hz s\textsuperscript{-1}. We also adopt $S_{\rm mean, 250} = 1.1\,$mJy at $250\,$MHz and $S_{\rm mean, 1400} = 0.046\,$mJy at $1400\,$MHz, approximately corresponding to the flux densities of the faintest known glitching pulsar in the Jodrell Bank Observatory (\ac{JBO}) Glitch Catalogue~\citep[PSR~B1911+11;][]
{Basu:2021pyd}.

Expected \ac{ToA} uncertainties for various SKAO's array configurations are derived in detail in \citet{Basu2025_SKA_EOS}. For a canonical pulsar near the zenith, SKA-Low can reach a timing precision of $\sim 2\,\mu$s after 4 min using 271 elements within 10 km (AA*), improving to $\sim 1\,\mu$s with 404 elements (AA4). These values are lower bounds: in practice, interstellar scatter broadening and pulse-to-pulse jitter will degrade the achievable precision, particularly at low frequencies. SKA-Mid can achieve high precision through longer integrations, benefiting from reduced ISM effects. Low-DM pulsars are therefore best suited for timing with SKA-Low, whereas SKA-Mid will be advantageous for precision mass and moment-of-inertia measurements.


Although the full array gives impressive sensitivity in terms of \acs{ToA} errors, a compromise on sensitivity is required to observe multiple sources simultaneously. Both SKA-Low and Mid will have the capability to form up to 16 concurrent (subarray) beams for pulsar timing in AA4. For SKA-Mid, 16 beams will also be possible for AA*, but SKA-Low will process up to 8 beams during AA*. This will allow a range of simultaneous observing possibilities as well as the allocation of smaller subarrays, and shorter integration times for brighter pulsars, to facilitate the efficient use of telescope time (e.g., \citealt{song:2021MNRAS}), whilst reducing the effects of pulse jitter. 


 In the case where a single array element is used, after a 4 minute integration of the pulsar described above, we can achieve $\sigma_{\rm ToA} = 0.5\,$ms for Low and $0.7\,$ms for Mid. However, given that the S/N scales with flux density, and in this case we chose a particularly faint pulsar ($\sim1\,$mJy at 250 MHz), a factor of $\sim$100 improvement in $\sigma_{\rm ToA}$ could be achieved when using single stations to target brighter ($\sim100\,$mJy) pulsars for a similar integration time. 


\subsection{Observing pulsar glitches with the SKAO}
\label{sec:glitchobs}

To probe the full glitch parameter space, an observing program should be devised in such a way that it resolves rotational variations on timescales from seconds to years for a large sample of pulsars. Moreover, as glitches can be seen as transient events, which may require alterations to observing methodology, a flexible approach to glitch detection and characterisation with the SKAO's observations is crucial. To optimally use the sensitivity of the SKAO's array to detect small glitches, in both real-time and retrospective searches (e.g., \citealt{melatos:2020ApJ}; \citealt{singha:2021MNRAS}), a high cadence is necessary to distinguish them from timing noise. 

The nominal cadence from routine timing programs is insufficient to capture some medium-term glitch recoveries on timescales below tens of days~\citep{Liu2024}. Resolving these recoveries will require either including the pulsar in a dedicated higher-cadence timing program~\citep{Basu:2019iam, lower:2021mnras, Basu:2021pyd, Liu2024} or employing a glitch detection approach such that cadences can dynamically be temporarily adjusted in order to resolve the recovery phase. 

High cadence observations can be made possible with both SKA-Mid and SKA-Low through flexible approach to scheduling. For instance, when a new glitch is detected during the routine timing program(s), a (subarray) beam can be temporarily allocated to the pulsar to enable improved cadence. In addition, observing cadences can be further enhanced by exploiting commensal observing capabilities, allowing access to many pulsars within the wide field-of-view, even if the main target of the observation is not a pulsar.  Such observations will be valuable for probing the prevalence of glitches and similar rotational features, as well as for discovering rare events such as
antiglitches (abrupt decreases in spin frequency primarily observed in magnetars \citep{Archibald2013}, atypical glitches, and glitches from unusual sources, such as millisecond or very old pulsars. Furthermore, it may be possible to alert other observatories of the occurrence of a glitch\footnote{Similarly, glitches detected by other observatories could be followed up using the SKAO's telescopes, where resources allow.} for rapid follow-up if SKAO's flexibility is limited at the time of the glitch event. The attainment of improved timing precision, close in time to a glitch epoch, will also be of interest to \ac{GW} detection facilities~\citep[e.g.,][]{Moragues2023}. 


Resolving the very short-term glitch effects requires a combination of high cadence and sufficiently long integration time to increase the likelihood of observing during and shortly after the event. The sensitivity and large field-of-view of the SKAO's arrays are ideal for catching glitches on the rise, revealing
the true prevalence of delayed spin-ups, and enabling studies of the short-term recovery process. Such an approach would also benefit from robust online glitch detection, triggering real-time alerts and diverting any available resources to the pulsar of interest as soon as possible. For instance, during long integration time observations of one of a selected group of bright, young pulsars which are part of a dedicated high-cadence monitoring programme, \acsp{ToA} can be calculated dynamically whenever a predefined S/N threshold is reached, resulting in multiple closely-spaced \acsp{ToA} from the same observation. In the event of a detected glitch, the observing schedule can be automatically re-prioritised to focus resources on the glitched pulsar, ensuring high-cadence coverage to capture the critical, rapid post-glitch evolution. The long integration times required could be achieved through the use of any idle SKA-Low station beams, in combination with commensal observations. This approach would maximise observing efficiency while maintaining the flexibility to respond dynamically to newly detected glitches. Targeted searches for possible magnetospheric emission changes associated with glitches---as recently seen in Vela~\citep{Palfreyman2018}---or for thermal X-ray signatures indicative of heat deposition, as expected if some glitches involve a pulsar crustquake~\citep{Bransgrove2020} will also be made possible by this approach. 

We evaluated the feasibility of detecting pulsar glitches with SKA-Low in near real time, following the approach used in the \ac{JBO} pulsar timing programme (See \citealt{Basu2025_SKA_EOS} for details). Simulations assumed fortnightly cadence, 4-min integrations, and ToA precisions of $1.8\mu$s (AA) and $1.1\mu$s (AA4). Glitches of amplitude $\Delta\nu = 1.8\times10^{-11}$Hz (the smallest recorded in the \ac{JBO} Glitch Catalogue) and $10\times$ larger were injected into white and red noise timing residuals. In the white noise case, AA4 provides a clear advantage, with mean detection significance exceeding $3\sigma$ for the smallest glitch. Red noise substantially reduces significance, and the improvement from AA* to AA4 becomes marginal. Simulations without glitches show similar high-sigma outliers, indicating potential false positives when red noise dominates. For PSR~B0410$+$69, the actual detection significance from the first post-glitch \ac{ToA} is $1.5\sigma$, below the threshold for confident real-time identification. These results indicate that AA4 can enable robust real-time detection of small glitches when red noise is low, a key advantage for millisecond pulsars, which typically exhibit little timing noise.
\begin{wrapfigure}{l}{0.5\textwidth} 
   \centering
   \vspace{-5pt}
\includegraphics[width=0.5\textwidth]{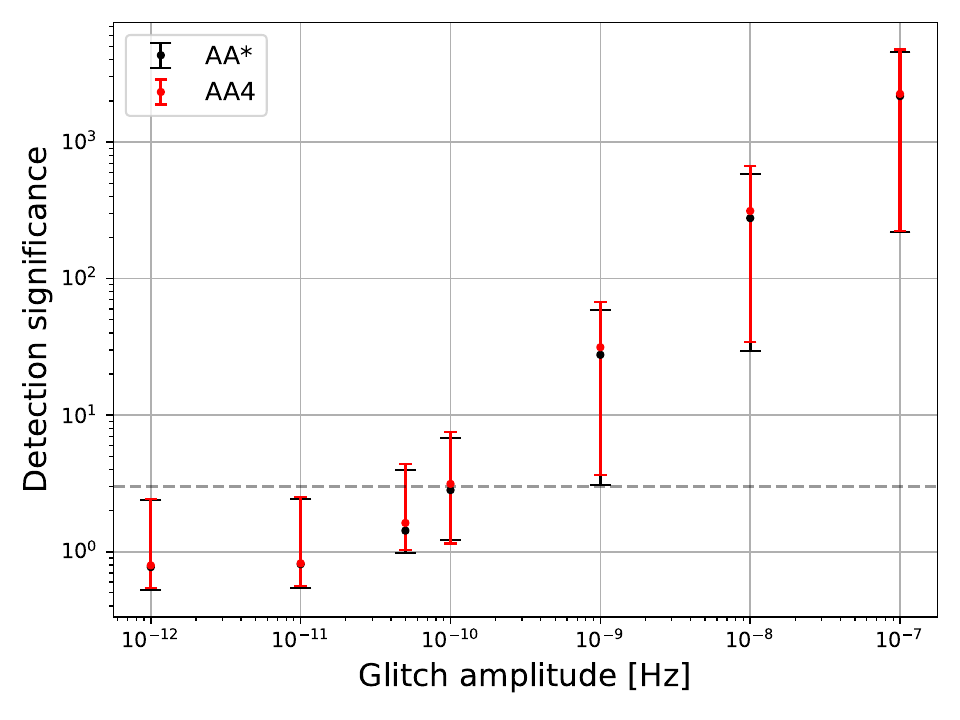}
\caption{Detection significance for a range of simulated glitch amplitudes in which red noise (according to the $A_{\rm red}$ and $\gamma$ values used above) is present in the data. Each point represents the median detection significance for 1000 realisations of the red noise for a given glitch amplitude. Error bars indicate the 68\% confidence interval, spanning the 16th to 84th percentiles of the distribution about the median. The horizontal dashed line denotes the $3\sigma$ level.}
\vspace{-10pt}
\label{fig:size_sig}
\end{wrapfigure}

Figure~\ref{fig:size_sig} shows the glitch detection significance that results from further simulations, all of which include red noise, over a wider range of glitch amplitudes. Glitches, occurring in a pulsar which exhibits red noise, are unambiguously detectable, using all SKA-Low stations within 10 km of the core, when their amplitudes exceed $\Delta \nu > 10^{-9}\,$Hz. While these simulations reflect a somewhat conservative scenario, i.e., by using a comparatively faint ($1.1\,$mJy) pulsar, they highlight that small glitches are still detectable with high confidence. For brighter pulsars, which are expected to yield smaller \acs{ToA} uncertainties, glitches of significantly lower amplitude would be detectable with even greater significance using SKA-Low. Such sensitive detections of glitches are crucial for disambiguating the effects of timing noise from glitches in pulsar timing residuals (e.g., \citealt{espinoza:2014mnras}) and providing constrains on the glitch mechanism.

The online glitch detection method relies solely on the most recent \ac{ToA}, assessing its deviation from a model based on preceding data. In principle, this enables real-time alerts when a \ac{ToA} arrives significantly early or late. 
However, the effectiveness of such alerts depends on the choice of pre-glitch interval and observing cadence. Long intervals increase the impact of red noise, which can mask small glitches, whereas short intervals reduce predictive power at low cadence.
High cadence can also be problematic if insufficient phase offset has accumulated by the first post-glitch epoch. Glitches below the real-time sensitivity threshold can still be identified retrospectively as phase discontinuities emerge with additional post-glitch data. Regular offline searches therefore remain essential to maintain a complete glitch sample~\citep[e.g.][]{espinoza:2014mnras}.


To maximise the efficient use of telescope time and avoid unnecessary follow-up, glitch detection systems should incorporate both timing residuals and pulse profile changes. Mode switching, intermittency, or radio-frequency interference (RFI) can all initially mimic glitches by introducing ToA discontinuities. Robust classification frameworks that jointly assess timing, profile variability and RFI are therefore essential for generating sensitive and reliable real-time alerts.


\subsection{Detecting free precession with the SKAO}

Free precession introduces a periodic modulation in pulsar timing residuals (see Section~\ref{sec:precession}), with a sinusoidal morphology superimposed upon the otherwise smooth spin-down evolution of the star. The period $P_{\rm mod}$ of this modulation may span timescales up to several years, depending on the \ac{NS}’s internal structure and ellipticity. These amplitude variations are, in principle, detectable provided they introduce residual fluctuations that exceed the average \acs{ToA} uncertainty. Moreover, the changing orientation of the pulsar beam with respect to the line of sight, caused by precession, may lead to periodic variations in the observed pulse profile shape, introducing additional phase shifts in the timing residuals, along with the observed polarisation position angle (e.g., \citealt{gao_etal_23, Desvignes+2024, Basu:2024qwt}), providing constraints on the geometric asymmetry. 

\begin{wrapfigure}{l}{0.5\textwidth} 
    \centering
    \vspace{-20pt}
\includegraphics[width=0.5\textwidth]{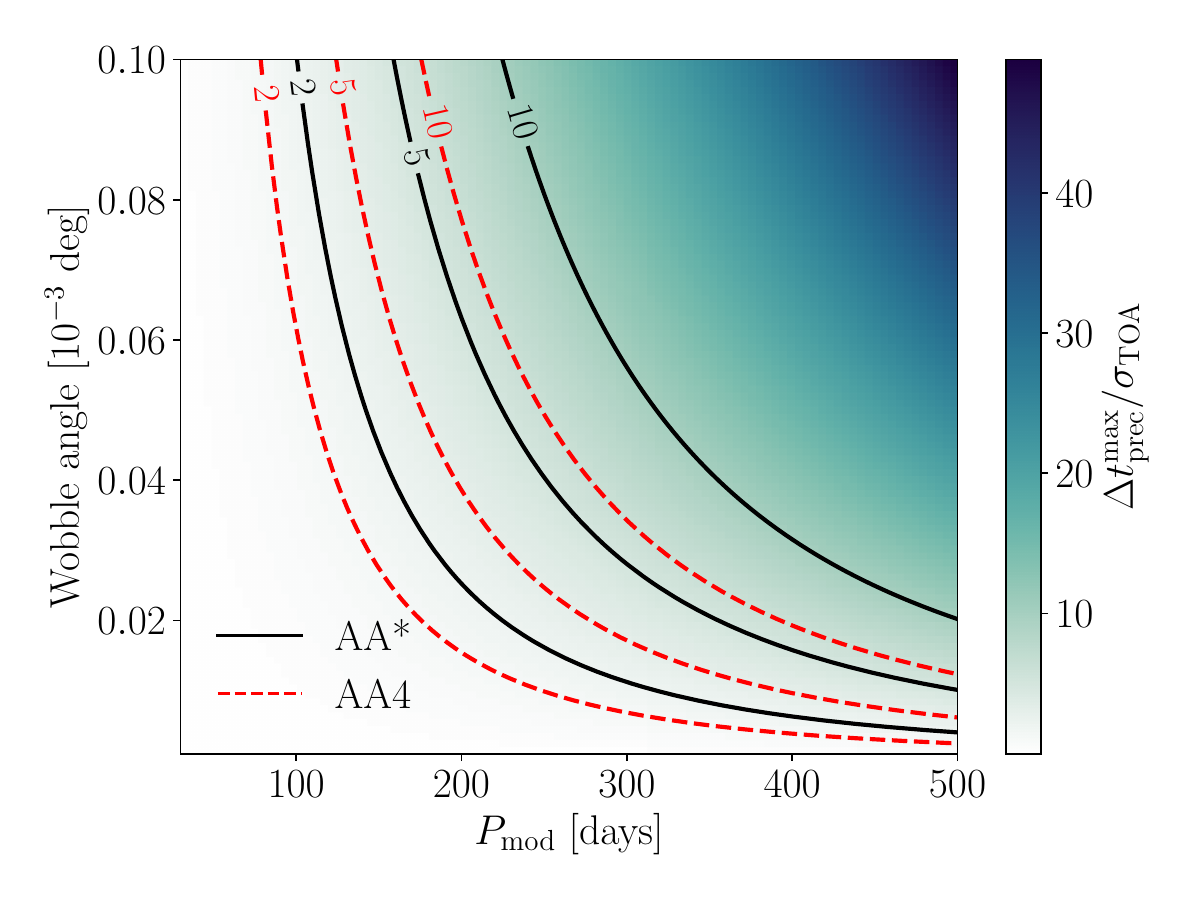}
\vspace{-20pt}
\caption{Contour map showing the significance of the effects of free precession on pulsar timing residuals for a range of wobble angles and precession periods for SKA-Low during AA* (black solid line) and AA4 (red dashed line). Darker regions indicate parameter combinations where the influence of precession exceeds the \acs{ToA} uncertainty, signifying a higher likelihood of detection. See Section~\ref{case_study} for the pulsar parameters used for this simulation.}
\label{fig:ska_precession}
\end{wrapfigure}
To examine the ability of the SKAO to resolve free precession in pulsar timing data, we use timing residuals generated as part of the simulations described in Section~\ref{sec:glitchobs}, for each array assembly, assuming each \acs{ToA} was measured using all available SKA-Low stations, as for a given pulsar, a higher timing precision is achievable. For simplicity, we did not include the effects of red noise, or any glitches in this analysis. We then inject oscillations introduced by free precession into the residuals for a range of wobble angles, $\theta$, and precession periods, $P_{\rm mod}$ (see Equation (61) of \cite{Jones2001} and \cite{Basu2025_SKA_EOS}) for more details. 
We then generate sensitivity curves as a function of $P_{\rm mod}$ and the wobble angle $\theta$. These are shown in Figure~\ref{fig:ska_precession}.The black and red lines show contour intervals corresponding to the 2, 5, and 10$\sigma$ levels of detection (see \citealt{Basu2025_SKA_EOS}) significance for AA* and AA4, respectively.These curves demonstrate that, assuming timing noise is negligible, AA4 will enable the detection of free-precession with over 5$\sigma$ confidence even for extremely small wobble angles ($\sim 2 \times 10^{-5^\circ} $) and precession periods of around 150 days. 

\subsection{Synergies with multiwavelength and multi-messenger facilities}
\label{sec:synergies}
Mass, spin, and \ac{MoI} measurements with the SKAO's observations will tightly constrain the dense matter \ac{EoS}, complemented by X-ray and \ac{GW} observations. X-ray constraints rely primarily on pulse profile modelling (\ac{PPM}), which infers mass and radius from emission originating at hot polar caps of rotation-powered millisecond pulsars, shaped by relativistic effects. Ray-tracing models yield not only $M$ and $R$ but also polar cap properties~\citep[see][and references therein]{Bogdanov:2019qjb,Bogdanov:2021yip}. NICER observations~\citep{Gendreau16} have already produced precise constraints for PSRs J0030$+$0451~\citep{Riley:2019yda, Miller19, Vinciguerra:2023qxq}, J0740$+$6620~\citep{Salmi24a, Dittmann24}, and J0437$-$4715~\citep{Choudhury:2024xbk}, and weaker ones for J1231$-$1411~\citep{Salmi24b}. Continued NICER monitoring will improve these measurements and extend them to additional sources. \ac{PPM} is also being applied to accreting \acp{NS}~\citep{Poutanen03, Kini24, Dorsman25, Salmi25}, and future missions such as eXTP~\citep{Li:2025uaw} and NewAthena~\citep{Cruise:2024mgo} are expected to deliver even tighter constraints.

\begin{wrapfigure}{r}{0.5\textwidth}
\centering
\vspace{-30pt}
\includegraphics[width=0.5\textwidth]{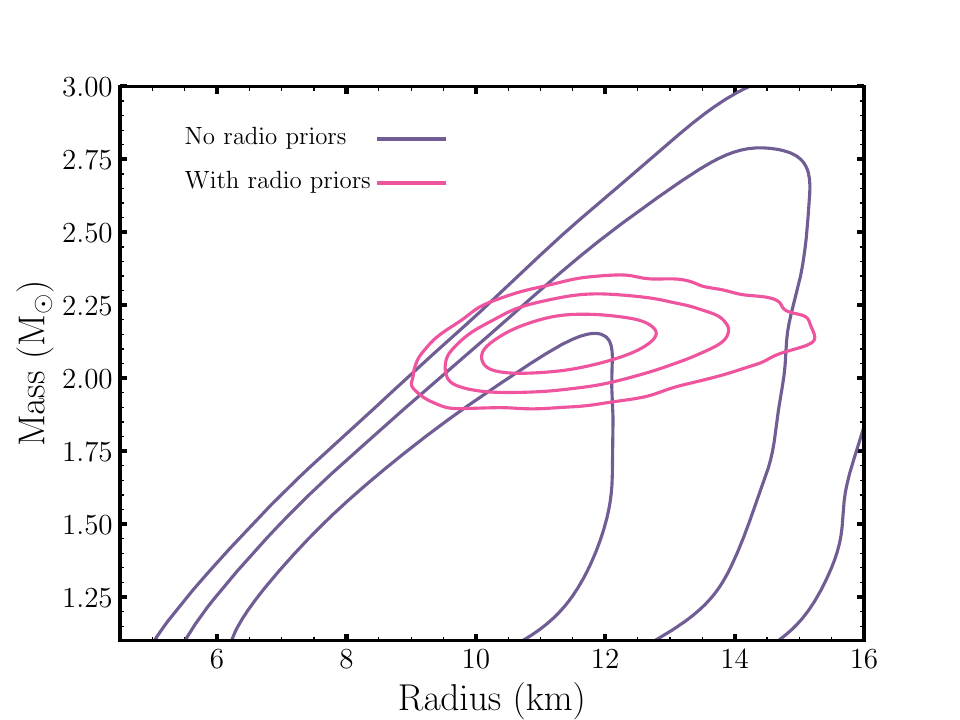}
\vspace{-15pt}
\caption{This figure, adapted from Figure 5 of~\citet{Riley:2021pdl}, illustrates the impact of the radio timing information (mass, distance and inclination) on the mass, $M$, and equatorial radius, $R$, inferred from pulse profile modelling of NICER data of the high-mass pulsar PSR~J0740$+$6620. It shows the two-dimensional marginal PDFs conditional on the informative radio priors from~\citet{Fonseca:2021wxt} (pink) and a diffuse uninformative prior in the absence of radio data (purple). The contours are the credible regions containing 68.3\% 95.4\% and 99.7\% of the posterior mass.  Without the radio timing information, the inferred mass-radius constraints are too broad to be useful for \ac{EoS} analysis.}
\label{fig:PPMnomass}
\end{wrapfigure}
In principle, the results from \ac{PPM} provide an independent check on the \ac{EoS} constraints delivered by SKAO's measurements but in practice the relationship is much closer. The NICER millisecond pulsars are also radio pulsars, and where available the radio-derived mass measurement (along with distance and inclination) is used as a prior in the \ac{PPM}. This was the case for PSR~J0740$+$6620~\citep{Fonseca:2021wxt} and PSR~J0437$-$4715~\citep{Reardon24}: In both cases, the mass measurement played a critical role in enabling the inference as illustrated in Figure~\ref{fig:PPMnomass}.
Improved mass priors provided by SKAO's observations for key \ac{PPM} sources will therefore be vital, highlighting the need for flexibility in timing those sources that are most promising for \ac{PPM} analysis. Information about the magnetospheric configuration (such as the magnetic and observer inclination angles;~\citealt{Oswald01.2026.SKA}), derived from radio observations with SKAO, is also expected to help reduce uncertainties relating to geometric/polar cap priors and model space involved in \ac{PPM}~\citep[see][for an example of a source where there are two potential mass-radius solutions associated with different magnetospheric geometries]{Vinciguerra:2023qxq}, resulting in tighter and more robust mass-radius inferences.

Looking further ahead, the feasibility of measuring the \ac{NS} radius of newly discovered pulsars by SKAO with X-ray \ac{PPM} not only depends strongly on the precise knowledge of the priors  outlined above, but also on the X-ray flux of the source (usually ranging between $10^{-13}-10^{-14}$\,erg~s$^{-1}$~cm$^{-2}$), the complexity of its surface temperature distribution, and the effective area and background knowledge of the available X-ray instrument. Consequently, the large range of parameters involved does not allow a firm identification of a distance limit below which we will be able to infer \ac{NS} radii for newly discovered sources by SKAO. However, especially with the future availability of NewAthena, X-ray observations significantly shorter than what is currently necessary with NICER will enable \ac{EoS} constraints (with a $<$3\% accuracy in the $M$-$R$ plane) for objects as distant as 1.5\,kpc~\citep{Cruise:2024mgo}.

\begin{wrapfigure}{r}{0.5\textwidth} 
    \centering
    \vspace{-20pt}
\includegraphics[width=0.5\textwidth]{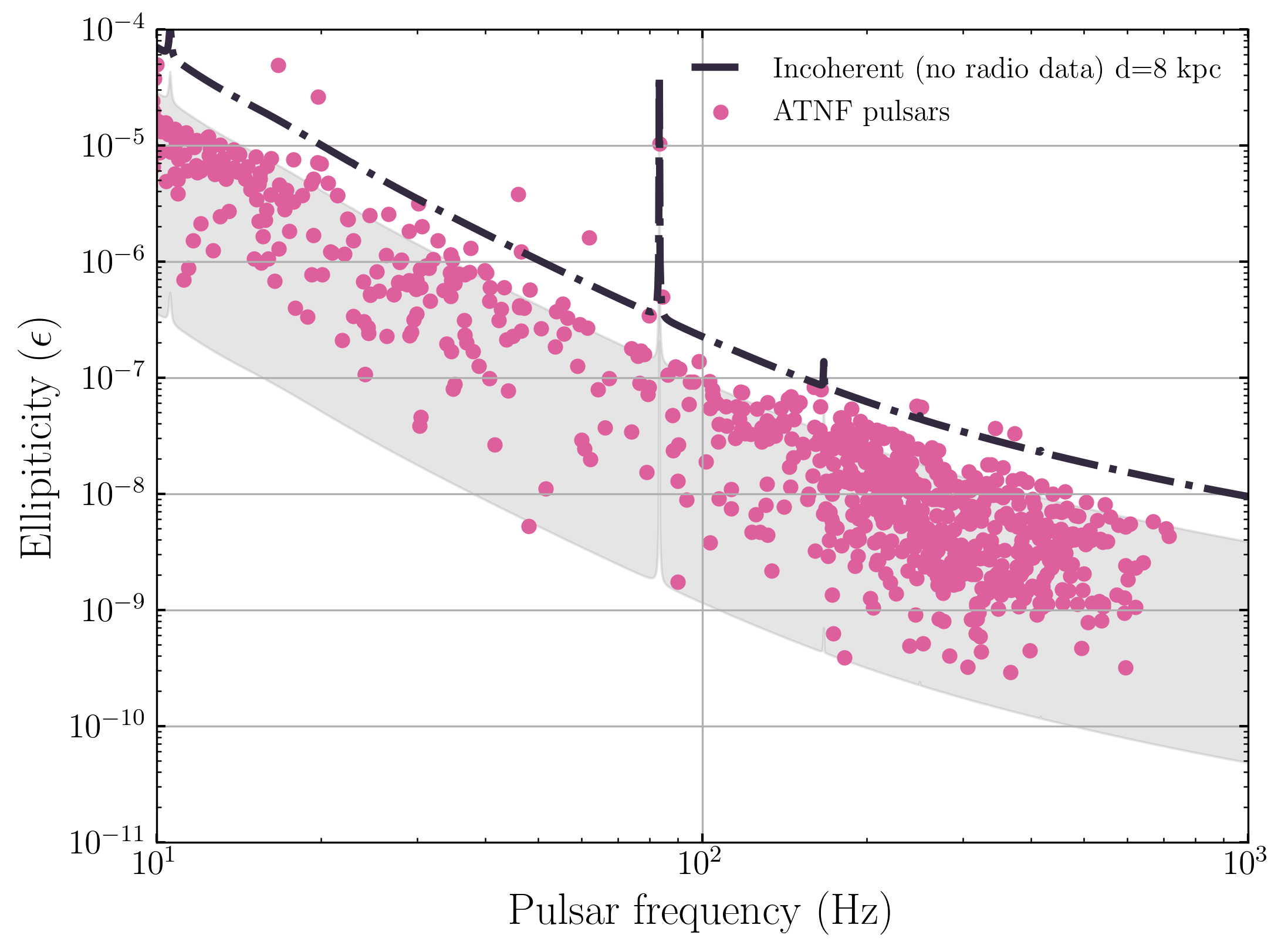}
\caption{Minimum detectable ellipticity for continuous \ac{GW} searches by ET. The black curve assumes a blind search with no radio data (i.e., an incoherent search with a total duration of 1 year with 10 days coherence time, as described in~\citealt{Branchesi:2023mws}) and an ellipticity limit calculated for a source at $8\,$kpc. Dots represent the minimum ellipticity detectable for pulsars in the ATNF Pulsar Catalogue \citep{manchester:2005atnf}, assuming that radio ephemeris enable a full 1 year coherent integration. The grey band highlights the region between $0.1\,$kpc and $8\,$kpc. Note that we plot ellipticity versus the rotation frequency of the pulsar itself, not the GW frequency, which will be twice the stars' rotation frequency.}
\vspace{-20pt}
\label{fig:ETpulsars}
\end{wrapfigure}

Electromagnetic observations complement \ac{GW} detections, jointly constraining \ac{EoS} systematics. This synergy was first demonstrated by the binary \ac{NS} merger GW170817~\citep{LIGOScientific:2017vwq} and its electromagnetic counterpart~\citep{LIGOScientific:2017ync}, where tidal deformation measurements ruled out the stiffest \acp{EoS}~\citep{LIGOScientific:2018hze, Chatziioannou:2024tjq}. Combining tidal properties with universal relations linking them to $I$ and the quadrupole moment~\citep{Yagi:2013awa} may further refine constraints on both. Next-generation detectors like the Einstein Telescope and Cosmic Explorer~\citep{Abac:2025saz, Evans:2021gyd} will detect numerous binary mergers, providing unprecedented insights into dense matter. Together with electromagnetic data, these \ac{GW} observations could test the presence of exotic states in the \ac{NS}s core, including anisotropic matter~\citep{Zuraiq:2023bpw} and twin stars~\citep{Christian:2021uhd}.

Complementary to terrestrial \ac{GW} observatories, the Laser Interferometer Space Antenna (LISA) is also expected to detect \acp{GW} emitted by a small population of `ultra-compact' ($P_{\rm b} \lesssim 10$\,min) relativistic double NS systems, that are otherwise undiscoverable by traditional radio pulsar surveys. A fraction of these pulsars will likely be detectable by the SKAO through the use of LISA-informed radio searches \citep{Kyutoku:2019MNRAS}. Timing observations of these systems with the SKAO would enable precision measurements of Lense-Thirring precession, providing additional independent constraints on the \ac{NS} \ac{MoI}, thus further informing the \ac{EoS}~\citep{Thrane:2020MNRAS}.


Beyond detecting mergers and ultra-compact binaries, future \ac{GW} interferometers may probe \ac{NS} interiors. As discussed, a \ac{NS} is not perfectly spherical but has an ellipticity $\epsilon$, producing continuous \ac{GW} emission if the deformation is non-axisymmetric. The emitted frequency $\nu_{\rm gw}$ corresponds to either $\nu_s$ or $2\nu_s$~\citep{Bonazzola96, Jaranowski98, Gittins24}. Microphysical modelling limits $\epsilon\lesssim10^{-6}$~\citep{Jones2001, Johnson2013, Gittins21a, Gittins21b}, though some \acp{EoS} allow larger deformations sustained by exotic or superconducting phases~\citep{Haskell07, Das2025}. Evolutionary modelling of millisecond pulsars in the $P$–$\dot{P}$ plane implies a residual $\epsilon\approx10^{-9}$, possibly due to a buried magnetic field~\citep{Woan18} in the supercounduction core. While such signals remain undetected, but current searches in LIGO-Virgo-KAGRA data are beginning to investigate astrophysically significant parameter spaces~\citep{HaskellBejger23}.

The next generation of ground-based detectors, particularly ET, will revolutionise \ac{GW} astronomy, potentially detecting continuous signals from hundreds of known pulsars if their ellipticities are at the higher end of the expected range, and a few tens of pulsars if these are closer to the lower limit~\citep{Abac:2025saz}. These estimates rely on pulsars with known radio ephemerides, which enable coherent searches over long data spans. Without ephemerides, blind searches must cover vast parameter spaces, reducing sensitivity~\citep{Riles23, Wette23} and limiting detections to nearby sources with $\epsilon \approx 10^{-7}$~\citep{Dergachev20, Branchesi:2023mws}. However, the SKAO’s discovery and precise timing of thousands of new pulsars~\citep{Keane01.2026.SKA, Levin01.2026.SKA} will greatly enhance the prospects of detecting continuous \acp{GW} with third-generation detectors, enabling the study of signals near the lower theoretical limits and offering complementary insights into the high-density \ac{EoS} beyond those probed by compact binary coalescences~\citep{Jones25}.

\section{Conclusions}
\label{sec:conclusions}

\acp{NS} offer access to the fundamental properties of matter at high densities, low temperatures, and large proton-neutron asymmetries---conditions that cannot be recreated on Earth. Their extreme environments make \acp{NS} unique laboratories for probing the nuclear \ac{EoS} and dense matter superfluidity. High-precision radio pulsar timing is a cornerstone of this endeavour, allowing us to infer global stellar properties like mass, \ac{MoI}, and spin frequency. When combined with observations in the X-rays, which measure the stellar radius, pulsar timing yields powerful constraints on the dense matter \ac{EoS}. Beyond global properties, pulsar timing also grants access to local phenomena, particularly glitches. These sudden spin-ups, now observed across many young pulsars, are the only observational window we currently have to study the dynamical properties of \ac{NS} superfluids.

The SKAO, especially once AA4 comes online, will be transformative for this field. Its unprecedented sensitivity and wide survey capabilities will allow us to time existing \ac{NS} systems with even higher precision and greatly expand the population of known pulsars, including new highly relativistic binaries and fast rotators. This will enable novel high-precision $M$ measurements and $I$ constraints with single-digit uncertainties, potentially pushing the limits of existing \ac{EoS} models. Future SKAO's observations will also open up a new regime in glitch science and potentially the detection of free precession, enabling systematic studies of superfluid dynamics across a diverse pulsar population.

To achieve these goals, the SKAO must operate in a range of observing modes. Large-scale surveys with SKA-Low and SKA-Mid will be essential for uncovering new systems with extreme and \ac{EoS}-relevant properties. Following these sources up with SKA-Mid for several hours with monthly cadence will be essential to build up the baselines relevant for the kinds of \ac{EoS} constraints outlined in this paper. To study glitches and other rotational irregularities, continuous monitoring over a wide range of timescales---from seconds to years---will be required. Given the transient nature of glitches and precession, these science cases also demand operational flexibility, including subarraying capabilities and a commensal timing mode, to capture and characterise events in real time.

Finally, synergies with other facilities, particularly X-ray observatories like NICER, and the upcoming NewAthena, and next-generation \ac{GW} facilities such as ET, will be essential to constraining subatomic physics. These multi-wavelength and multi-messenger approaches will not only provide deeper insights into the properties of ultra-dense matter but also open the door to probing physics beyond the Standard Model, from \ac{DM} interactions to alternative theories of gravity.


\section*{Acknowledgments}
Pulsar research at Jodrell Bank is supported by a consolidated grant (ST/T000414/1 and ST/X001229/1) from the UK Science and Technology Facilities Council (STFC).
V.~G. is supported by a UKRI Future Leaders Fellowship (grant number MR/Y018257/1). M.~E.~L. is supported by an Australian Research Council (ARC) Discovery Early Career Research Award DE250100508.
P.~C. acknowledges the support from the European Union's HORIZON MSCA-2022-PF-01-01 Programme under Grant Agreement No. 101109652, project ProMatEx-NS. This project has received funding from the European Union’s Horizon 2020 research and innovation programme under the Marie Skłodowska-Curie grant agreement No. 101034371. 
H.~H. acknowledges support from the PRIME programme of the German Academic Exchange Service (DAAD) with funds from the German Federal Ministry of Education and Research (BMBF). 
N.R. is supported by the European Research Council (ERC CoG No. 817661 and ERC PoC No. 101189496), and grants SGR2021-01269, ID2023-153099NA-I00, and CEX2020-001058-M.

\bibliographystyle{abbrvnat-maxbibnames4}
\bibliography{bibliography_SKAO.bib} 

\end{document}